\documentstyle[pre,preprint,aps,epsf]{revtex}
\draft
\tighten

\begin{document}

\author{Yuriy E. Kuzovlev \\
%EndAName
Donetsk Physics and Technology Institute of NASU\\
340114, Donetsk, Ukraine }
\title{Kinetical theory beyond conventional approximations and 1/f-noise }
\date{March 21, 1999 }
\maketitle

\begin{abstract}
The explanation of existence and ubiquity of 1/f-noise is under
consideration based on the idea that 1/f-noise has no relation to long
lasting processes but originates from the same mechanisms what are
responsible for the loss of correlations with the past, shot noise and fast
relaxation. According to this idea, in a system which produces incident
events being indifferent to their totally happened amount the fluctuations
in the production of events grow proportionally to its average value. Such a
free memoryless behaviour results in the 1/f  long-living statistical
correlations without actual long-living causality.

The phenomenological theory of memoryless random flows of events and of
related Brownian motion is presented which closely connects 1/f-noise and
non-Gaussian long-range statistics, both expressed in terms of only
short-range characteristic scales.

The exact relations between 1/f-noise and equilibrium four-point cumulants
in thermodynamical systems are analysed. The presence of long-living
four-point correlations and flicker noise in the Kac's ring model is
demonstrated.

The problems of kinetical approaches to noise and irreversibility in
Hamiltonian systems are touched, and the general idea is confirmed in the
case of gas.

It is argued that correct construction of gas kinetics in terms of
Boltzmannian collision operators needs in the ansatz whose meaning is
conservation of particles and probabilities at the path from in-state to
out-state inside the collision region. This reason results in the infinite
set of kinetical equations which describe the evolution of many-particle
probability distributions on the hypersurfaces corresponding to encounters
and collisions of particles. These equations reduce to usual Boltzmann
equation only in the spatially uniform case, but in general forbid molecular
chaos because of spatial correllations of colliding particles.

The formulated kinetics are applied to investigate the statistics of
self-diffusion in equilibrium gas. The peculiar behaviour of the four-order
cumulant of Brownian displacement of a gas particle is found being identical
to 1/f-fluctuations of diffusivity and mobility. This is the example of
dynamical system which produces no slow processes but produces 1/f-noise.
\end{abstract}

\thinspace \thinspace \thinspace

{\bf CONTENTS}

1. Introduction.

2. The problem of 1/f-noise (flicker noise).

3. Phenomenology and philosophy of 1/f-noise.

4. Brownian motion revised.

5. Dynamics and 1/f-noise.

6. Generalized fluctuation-dissipation relations.

7. Current state and paradoxes of gas kinetics.

8. Correct reformulation of Boltzmannian kinetics.

9. Self-diffusion and diffusivity 1/f-noise.

10. Resume.

\section{INTRODUCTION}

The low-frequency 1/f-noise (flicker noise) was discovered by Johnson in
1925 in electronic lamps. At present 1/f-noise is known in giant variety of
systems and processes in physics, chemistry, biology, medicine, cosmic and
Earth sciences and in human activity. Perhaply, thousands of experiments
were performed during many years in order to comprehend this apparently
mysterious phenomenon. Howevere, a conventional theory of 1/f-noise is still
lacked.

It seems that the situation testifies the absence of true ideas. May be,
that is why there are no new substancial reviews on 1/f-noise like ones
[1,2] published in 1981. As before, usually 1/f-noise is believed be
produced by some slow processes with a broad distribution of long relaxation
times (life-times), and 1/f spectrum is thought as superposition of
Lorentzians. But in most particular cases the hypothetical slow processes
are not identified.

The popular idea about long life-times is evoked by common opinion that any
long-range statistical correlations associated with 1/f spectrum reflect
some real long-living causal connections between the events which took place
in the past and which could occur in the future. But as long ago as in 1950
N.S.Krylov [3] argued the necessity to make difference between statistical
correlations and factual causality.

In view of this distinction, the non-trivial explanation of 1/f-noise is
possible [4-11] attributing its origin just to absence of long-living
causality. More concretely, to processes which currently forget their own
history and thus could not keep an eye on its characteristics. The flow of
incident events produced by such a process have no certain ''number of
events per unit time'', and just this circumstance implies 1/f-noise.
Indeed, if a process remembers nothing about old events and thus about
previous number of events per unit time, how could it know about the future
one? If the past is of no importance, all the more the future is too.

From this viewpoint, the long-range 1/f correlations are only the
manifestation of a freedom of the random process. Correspondingly, the
shorter is life-time of real causal memory the larger is magnitude of 1/f
fluctuations [4-8], i.e. just the inverse rule takes place, and only if the
number of events per unit time is somehow anticipated by underlying dynamics
one gets a process without 1/f-noise.

For a first look, this picture is in contradiction with traditional hopes
that absence of long-living memory should result in white noise, Marcovian
behaviour and Poissonian flows of events. But in fact the latter is only
very special variant of pure randomness. One has not to break foundations,
merely the central limit theorem is not valid, and so another chapters of
the probability theory [12] and theory of randomness in dynamical systems
[13] should be involved.

The main purpose of this maniscript is to review our own works on this
subject because their English translations are hardly available. We do not
try to review either the infinity of experimental data or recently suggested
models of 1/f-noise (although some models may be in fact more close to our
ideology than to conventional tendency to long life-times). Instead, we
would like to show that if one avoids the limitations and approximations
standardly attracted when constructing models of noise then both the
phenomenology [4-8,10] and approaches based on statistical mechanics [9,11]
naturally predict 1/f contribution to whole noise without long life-times
and slow processes.

In particular, the correct procedure of derivation of gas kinetics from
exact BBGKY equations demonstrates [9] that 1/f-noise exists even in gas in
the Boltzmann-Grad limit, that is in the system where nothing like slow
processes can exist.

Unfortunately, up to now those who deal with rigorous statistical mechanics
payed almost no attention to 1/f-noise and likely had no suspicions that its
nature could be as fundamental as that of relaxation and irreversibility in
Hamiltonian systems. In our opinion, 1/f-noise is typical companion of
relaxation and dissipation phenomena in time-reversible dynamics, and its
sense is the absence of absolutely certain (well time-averagable)
quantitative rates of relaxation and irreversibility.

\section{THE PROBLEM OF 1/F-NOISE}

\subsection{FORMAL DEFINITION OF 1/F-NOISE}

Let $X(t)$, $\left\langle X\right\rangle $ and $\delta X(t)\equiv
(X(t)-\left\langle X\right\rangle )/\left\langle X\right\rangle $ be a
random process, its mean value and its relative variation, respectively. The
process will be called 1/f-noise if the power spectrum $S_{\delta X}(f)$ of
$\delta X(t)$ looks approximately as $1/f$ at low frequencies and has no
tendency to saturation at $f\rightarrow 0$ . At high frequences, in practice
$S_{\delta X}(f)\rightarrow S_\infty \neq 0$ (the white noise level). The
brackets $\left\langle {}\right\rangle $ denote the ensemble averaging.
Sometimes spectra like $S_{\delta X}(f)\propto f^{-\gamma }$ are observed
with $\gamma $ noticably different from unit, but we shall not discuss this
unprincipal difference, except the meaning of formal non-stationarity at
$\gamma >1$.

The distinctive temporal property of 1/f-noise is the impossibility of
precise measuring its mean value $\left\langle X\right\rangle $ with the
help of time smoothing procedure. Indeed, the relative deviation

\[
\delta x(T)\equiv [\frac 1T\int_0^TX(t)dt-\left\langle X\right\rangle
]/\left\langle X\right\rangle
\]
of the time-averaged value from $\left\langle X\right\rangle $ has the
variance
\[
\left\langle \delta x^2(T)\right\rangle \approx \int_{f_0}^\infty [\sin (\pi
fT)/\pi fT]^2S_{\delta X}(f)df
\]
with $f_0$ being the inverse duration of observations. When $T$ increases,
first this integral decreases as $\left\langle \delta x^2(T)\right\rangle
\approx S_\infty /T$ , in accordance with the large number law, but later
turns into nearly constant.

Formally, $\delta x(T)$ could either decrease in an extremely slow
logarithmic way or even slowly grow if $\gamma >1$ . But practically any
case results in existence of nonzero limit of $\sqrt{\left\langle \delta
x^2(T)\right\rangle }$ at large intervals $T$ . Therefore, some
characteristical uncertainty of the mean value of $X(t)$ does exist, the
so-called 1/f-floor or flicker floor.

\subsection{EXAMPLES OF THE FLICKER FLOOR}

a) Let $X(t)$ be the number of alpha-particle decays per unit time in a
piece of radioactive matter (for instance, in $Am$ with life-time of order
of 1000 years). The experimentally discovered 1/f fluctuations of the rate
of decays give the evidence that the decay probability possesses relative
uncertainty $\sim 10^{-4}$ .

b) Let $X(t)$ be the electron drift velocity in a current-carrying slightly
doped semiconductor. In accordance with the empirical Hooge's formula,
$S_{\delta X}(f)\approx \frac \alpha f$ , where $\alpha \approx 2\cdot
10^{-3} $ , at frequences lower than tens kHz [1,2]. The corresponding
relative uncertainty of drift velocity and mobility is of order of $0.1$ ,
i.e. surprisingly large.

c) The giant relative uncertainty, $\sim 1$ or even larger, characterizes
1/f-fluctuations of the inner friction and dissipation rate in excited
quartz crystals [14,15]. In this case $X(t)$ can denote the energy
dissipated per unit time.

\subsection{IT IS EVERYWHERE! MORE EXAMPLES}

i) Different solid, liquid and gaseous electric conductors [1,2]. $X(t)$ is
the charge transported per unit time, or resistance. Many of experiments are
well described by $S_{\delta X}(f)\approx \alpha /nf$ , where $\alpha
=10^{-7}-10^1$ for various media and $n$ is the number of charge carriers in
a sample under observation.

ii) Highway traffic. $X(t)$ is the number of cars passing per unit time.

iii) Laser light Raman scattering [16]. $X(t)$ is the number of photons
scattered per unit time.

iv) Chemical reactions in small volumes. $X(t)$ is the number of reaction
events per unit time.

v) Definite type of diabetics. $X(t)$ is the consumption of insuline per
unit time.

vi) Cosmic rays. $X(t)$ is the current intensity of cosmic radiation.

vii) Music. $X(t)$ is the volume or dominating tone or other property of a
sound track.

viii) Earth's rotation. $X(t)$ is the angular velocity of rotation.

ix) DNA. $X(t)$ is the number of definite complementary pairs per unit
length of DNA sequence. In this case the distance along the sequence serves
instead of time (for review and some original ideas see [17,18]).

x) Thermal equilibrium voltage noise of a currentless disconnected
conductor. $X(t)$ is the instantaneous power spectrum $S$ of the noise (that
is the spectrum currently obtained by filtering, squaring and smoothing over
a finite time interval). Voss and Clarke [19] supposed that the resistance
1/f-noise in nonequlibrium current-carrying conductors (the example i) )
means the existence of similar 1/f-fluctuations of the power spectrum $S$ of
equilibrium white noise, in accordance with Nyquist formula if extended to
slowly varying resistivity, and experimentally confirmed their hypothesis.

\subsection{1/F-NOISE OF EQUILIBRIUM WHITE NOISE\\
AND HIGHER-ORDER CUMULANTS}

The example x) is of special importance for us, because it demonstrates that
1/f-noise in disturbed thermodynamical systems can be theoretically
investigated with the help of four-point equilibrium correlators. Indeed,
the spectrum $S$ of equilibrium noise $u(t)$ is measured as a quadratic
functional of $u(t)$ , so the correlation function and spectrum of
fluctuations of $S$ can be expressed with four-point correlators of $u(t) $
[20].

To be more precise, the information about 1/f-noise is hidden in those
specific part of four-point correlators which can not be reduced to lower
order correlations, i.e. in the four-point cumulants (semi-invariants)
$\left\langle u_1,u_2,u_3,u_4\right\rangle $ defined by
\[
\left\langle u_1u_2u_3u_4\right\rangle =\left\langle u_1u_2\right\rangle
\left\langle u_3u_4\right\rangle +\left\langle u_1u_3\right\rangle
\left\langle u_2u_4\right\rangle +\left\langle u_1u_4\right\rangle
\left\langle u_2u_3\right\rangle +\left\langle u_1,u_2,u_3,u_4\right\rangle
\]
where $u_i\equiv u(t_i)$ and the equaliy $\left\langle u(t)\right\rangle =0$
is supposed. In most practically interesting cases the equilibrium noise
$u(t)$ looks as white noise with a single characteristic relaxation time
$\tau _c$ which determines its average spectrum,
$\left\langle S\right\rangle=\int \left\langle u(\tau )u(0)\right\rangle d\tau \approx
2\left\langle u^2\right\rangle \tau _c$ . The first three terms on the right-hand side
reflect only this short-range relaxation and say nothing about low-frequency
fluctuations of $S$ described by the last cumulant term.

If the equilibrium noise $u(t)$ was purely Gaussian, this term would turn to
zero. {\it So the existence of 1/f-noise of white noise inevitably means
that }$u(t)${\it \ possesses non-Gaussian statistics.}

\subsection{WHAT IS THE QUESTION?}

Though being everywhere the 1/f-noise phenomenon still stays without a
commonly accepted explanation. Therefore the natural question arises: could
it be explained in some general manner or not?

At present, the most popular approaches to this issue are based on the idea
that 1/f-noise is caused by some ''slow processes'' having large life-times
[1,2]. For instance, in case of electric 1/f-noise in solids the slow
processes could be associated with scattering of electrons by metastable
movable structural defects or with their trapping by long-living structural
inhomogenities. With no doubts, some part of low-frequency noise in solids
is produced by such mechanisms and reflects real slow processes. It is
interesting that formal summation of Lorentzians corresponding to a
distribution of activation energies can reproduce factually observed
connections between frequency and temperature dependencies of 1/f-noise in
metals [1].

But, at the same time, there are no doubts that this aproach is unsufficient
for complete explanation of 1/f-noise even in solids, not to mention the
electric 1/f-noise in liquid metals and other liquids, all the more, various
1/f-noises as in the above examples. May me, Tandon and Bilger [21] were
right that the nice idea about summation of Lorentzians perhaply is a
disaster for building a true 1/f-noise theory.

\section{PHENOMENOLOGY AND PHILOSOPHY\\ OF 1/F-NOISE}

\subsection{1/F WITHOUT SLOWNESS}

The non-standard idea we are carrying from 1982 [4-11] is that 1/f-noise in
general has no relation to actual (physical, chemical, etc.) slow processes,
long-living memory and causal correlations, but, in opposite, manifests the
absence of thats.

To feel the idea let us consider the example v) . Assume that after taking a
doze of insuline a diabetics completely restores the basic normal state of
his organism and takes next doze only under the necessity, not under a
time-table. By definition, the basic state does not depend on how frequently
he taked the drag during last day, last week, etc., all the more during last
year. The past is constantly forgotten. If it is so, then the organism has
neither reasons nor ways to establish some definite ''doze per unit time''.
And one has no grounds to hope that the time avering will bring him some
predictably certain result, just because of absence of long-living memory!
{\it The events are accumulated following the principle ''let it be what
occurs to be''}. The future is as unknown as the past.

But such a free behaviour, without certain limit of the time smoothed doze,
means the existence of 1/f-fluctuations of ''doze per unit time''. Why 1/f ,
not other? If the smoothed doze had some definite limit but dependent on
occasion, the low-frequency spectrum of the doze would acquire the
delta-function $\delta (f)$ . In absence of the limit the spectrum should
have the same dimensionality on the frequency scale, that is be nothing but
$1/f$ [4].

By its nature, the 1/f-noise produced in such a way has no saturation at
zero frequency, and one can not saturate it by a noisy perturbation.

\subsection{STATISTICAL LONG CORRELATIONS\\ WITHOUT LONG-LIVING CAUSALITY}

The existence of the 1/f floor means that in a random flow of events the
probable random deviation of the number of events from its ensemble-averaged
value grows nearly as the average, the deviation is approximately
proportional to time. The loss of information about old events gives the
natural ground for this property : {\it merely the system can not
distinguish what part of happened events could be qualified as a proper part
and what as a deviation from it}. Any present result equally may serve as
the starting point for further creation of events.

{\it The remarkable paradox is that this memoryless behavior can not be
statistically described somehow except the language of long-living
correlations!}

All the events which belong to one and the same time piece seem mutually
correlated, because all thats equally participate in the result. But there
are no actual long-living causality beyond the corresponding statistical
correlations [4-10].

Thus slow processes and slow statistical correlations are not one and the
same thing. This important circumsance was underlined by N.Krylov [3] in
1950, in the frame of his critical revision of statistical thermodynamics.
But he pointed out no examples. We supposed in 1982 that 1/f-noise is just
the case [4].

\subsection{SELF-FORGETTING FLOWS OF EVENTS\\ AND CAUCHY STATISTICS}

Let us consider a stationary random process of accumulation of some quantity
$Q(t)$ , with instantaneous speed $u(t)$ and $\Delta Q(t)\equiv
Q(t)-Q(0)=\int_0^tu(t^{\prime })dt^{\prime }$ being the increment during
time interval $(0,t)$ , $t>0$ . For instance, $\Delta Q$ may be an amount of
drag consumpted by diabetics. In other example $\Delta Q$ is spatial
displacement of Brownian particle subjected to a constant force $x$ [6]. In
this case we take in mind so large $t$ that the drift component of
displacement much exceeds its diffusional component,
that is $U_0t>>\sqrt{2D_0t}$
(here $U_0$ , $D_0$ and $\mu =D_0/T$ are the average drift velocity,
diffusivity and mobility, defined by the relation $U_0\equiv \left\langle
u(t)\right\rangle =\mu x$, with $u(t)$ and $T$ being velocity and
temperature, respectively). One more example is the charge transport in
electric junction, if applied voltage much exceeds $T/e$ and so almost all
electrons jump in one and the same direction. Then the number of transported
electrons $\Delta Q(t)$ also can be treated as a random flow of events
(random walk) with definitely (let positively) directed steps, $\Delta
Q(t)\geq 0$ .

Let $\tau _0$ be the maximum of time scales characterizing causal
correlations of $u(t)$ , i.e. the life-time of dependence of the process on
its history. We want to analyse the asymptotical statistics of $\Delta Q(t)$
and of the time-averaged speed (rate) of the walk $U\equiv \Delta Q/t$ (doze
per unit time, drift velocity, current, etc.) at $t>>\tau _0$ , when all the
real memory effects should disappear [6,8].

Consider the amounts accumulated during two successive time intervals each
longer than $\tau _0$ , $\Delta Q(t)\equiv Q(t)-Q(0)$ and $\Delta Q^{\prime
}(t)\equiv Q(2t)-Q(t)$ , the summary amount $\Delta Q(2t)=\Delta Q(t)+\Delta
Q^{\prime }(t)$ and the corresponding time-smoothed rates $U(t)\equiv \Delta
Q(t)/t$ , $U^{\prime }(t)\equiv \Delta Q^{\prime }(t)/t$ and $U(2t)\equiv
\Delta Q(2t)/2t=[U(t)+U^{\prime }(t)]/2$ .

Because of $t>>\tau _0$ we believe that $U$ and $U^{\prime }$ are
independent random values. From the other hand, at $t>>\tau _0$ the result
should be insensitive to $t$ , because the system forgets how long ago the
measurement had started and how large is $t$ as compared with $\tau _0$ .
Consequently , $U(2t)$ should coinside with $U(t)$ in the statistical sense
[6], i.e. $2U(2t)=U(t)+U^{\prime }(t)$ is equivalent to $2U(t)=U(t)+U(t)$ .
We come to conclusion that {\it the sum }$U+U^{\prime }${\it \ of two
independent random values behaves as the sum }$U+U${\it \ of two completely
dependent values}! In other words, $\Delta Q(2t)$ does not statistically
differ from $2\Delta Q(t)$ .

That is nothing but the distinctive feature of Cauchy statistics. It is easy
to verify this in case of a random value $z$ subjected to symmetrical Cauchy
distribution whose probability density is $W(z,w)=\frac w\pi (w^2+z^2)^{-1}$
, with $w$ being the width of distribution. Indeed, the convolution of
$W(z,w)$ with itself equals to $W(z,2w)$ which is nothing but the probability
density of the doubled variable $2z$ .

The same is true also with respect to maximally asymmetrical Cauchy
distribution [12] whose characteristic function (CF), $\int \exp
(ikz)W(z,w)dz$ , looks as $\exp [-ikw\ln (-ikcw)]$ , where the width
parameter $w$ determines the magnitude of random value, $z\propto w$ , and
$c\sim 1$ .

Obviously, if $z$ is identified with $\Delta Q$ then the width $w$ must be
proportional to time, $w\propto t$ . Hence, the statistical equivalence of
$U(2t)$ and $U(t)$ determines both the probabilistic and temporal properties
of $\Delta Q(t)$ .

Because Cauchy statistics makes no difference between dependent and
independent contributions to a random walk, it is the naturally suitable
statistics to describe correlations what have no actual causality beyond
them. Its characteristical temporal property is that the magnitude of
accumulated fluctuations is proportional to time, $\Delta Q(t)\propto t$
(that is the central limit theorem fails).

\subsection{CAUCHY STATISTICS AND LONG CORRELATIONS}

Now we come to important point. The formal defect of the ideal Cauchy
statistics is the infinity of statistical moments. Indeed, the identity of
$U+U^{\prime }$ and $U+U$ in the sense of moments is possible only if the
variance of $U$ is infinite. This means that probability density has a long
power-law tail.

In practice one hopes that random rates are somehow bounded and have finite
cumulants and moments. From the other hand, even if a statistical ensemble
has infinite moments, this infinity never could be observed, at least while
finite time intervals. Any moment practically determined with the help of
time averaging always is finite with unit probability.

But no theory can do without ensemble language. Therefore, we need in an
ensemble which would be close to the Cauchy ensemble but possess finite
moments and be applicable to real time-limited observations.

Clearly, such an ensemble will allow for as large $\Delta Q(t)$ deviations
from its most probable behaviour as long is the observation time, in
accordance with the temporal scaling, $\Delta Q(t)\propto t$ , of the ideal
Cauchy random walk. Hence, this ensemble inevitably includes long-living
statistical correlations although thats express nothing mote than the
perfect randomness of the process.

\subsection{SELF-FORGETTING AND SCALE INVARIANCE}

The universal features of such statistics can be deduced from the simple
requirements to follow [6,8].

a). The ensemble average value of the accumulated amount is strictly
proportional to time, $\left\langle \Delta Q(t)\right\rangle =U_0t$ , that
is the mean value of time-averaged rate does not depend on $t$ ,
$\left\langle U(t)\right\rangle =U_0$ .

b). In the limit $t/\tau _0$ $\rightarrow \infty $ , the quantities $\Delta Q
$ and $U$ are constituted by many causally independent incidents, so the
probability distribution of $\Delta Q$ should acquire an infinitely
divisible form. In accordance with the general Levy-Khinchin representation
of infinitely divisible distributions (for one-directed asymmetrical walks
with $\Delta Q\geq 0$ ) [12], the CF of $\Delta Q$ can be written as
\[
\left\langle \exp \{ik\Delta Q(t)\}\right\rangle =\exp \{tU_0\int_0^\infty
(e^{ikq}-1)G(q,t)\frac{dq}q\}
\]
The essence of this formula is that function $G$ is non-negative,
$G(q,t)\geq 0$ . Besides, because of condition a) , the equality
$\int_0^\infty G(q,t)dq=1$ should be satisfied.

c). Due to $t>>\tau _0$ , $\Delta Q(t)$ is believed to asymptotically behave
as a random walk with independent increments. This means that logarithm of
CF is approximatelly proportional to time, $\ln \left\langle \exp \{ik\Delta
Q\}\right\rangle \propto t$ , i.e. $G(q,t)$ becomes almost time independent
at large $t$ .

d). $\Delta Q(t)$ should asymptotically behave in a scale-invariant way,
because of indifference to the short-time scale $\tau _0$ and absence of
larger characteristic scales. The concrete form of scale invariance is
prompted by the mean law $\left\langle \Delta Q\right\rangle \propto t$ and
looks as
\[
\Delta Q(\lambda t)\sim \lambda \Delta Q(t)
\]
where symbol $\sim $ denotes the approximate statistical equivalence. Thus
though being random $\Delta Q(t)$ grows nearly proportionally to time.

The physical meaning of this assertion was already discussed: a process what
is currently forgetting its previous accumulations (number of events,
Brownian displacement, etc.) is unable to make difference between its mean
part and deviation part. {\it Any result of previous behavior is equally
acceptable, regardless of its difference from ensemble expectations!}
Therefore it is more correct to say that just the average $\left\langle
\Delta Q(t)\right\rangle \propto t$ reflects the scaling $\Delta Q(t)\propto
t$ of random increments. That is why we must not especially carry
out $ik\left\langle \Delta Q\right\rangle $ from $\ln \left\langle \exp
\{ik\Delta Q\}\right\rangle $ .

e). In the limit $t/\tau _0$ $\rightarrow \infty $ the time-averaged rate
tends to $U_0$ . Thus we even want the process be ergodic. We will see that
the ergodicity automatically follows from a)-d) , but it is satisfied only
in a logarithmically weak sense, as well as c) and d) do.

\subsection{CHARACTERISTIC FUNCTION OF\\ PURELY RANDOM FLOW OF EVENTS}

If being formally punctual the scale-invariance condition d) would mean that
$\lambda G(q,\lambda t)=G(q/\lambda ,t)\,$, that is $\,G(q,t)=q^{-1}\Phi
(q/Vt)$ . Here $\Phi $ is formally arbitrary function and $V$ is some
characteristical rate introduced to make the argument $q/Vt$ dimensionless.
But this purely invariant form is incompatible with conditions b) and c).
Indeed, if $\Phi (0)\neq 0$ then the integration diverges at $q\rightarrow 0$
, while if $\Phi (0)=0$ then the condition c) will not be satisfied.

Hence we should allow for a slight violation of the scale invariance. We can
expect that necessary violation will be really weak because the divergency
is logarithmically weak.

The correction must cut the divergency at small values of $q$ , $q<Q_0$ ,
with $Q_0$ being some characteristical ''microscopic''
scale for measuring $\Delta Q$ . Such formal correction reflects
a really inevitable violation of
the invariance at small increments $<Q_0$ as well as at small time
intervals. For instance, one could simply replace the integration from zero
to infinity by integration starting at $q=Q_0$ (there are many other
possibilities but results.will be practically the same). Then condition a)
yields
\[
1=\int_{Q_0}^\infty \frac 1q\Phi (\frac q{Vt})dq\approx \Phi (Q_0/Vt)\ln
(\frac{Vt}{Q_0})\approx \Phi (0)\ln (\frac t{\tau _0})
\]
where $\tau _0=Q_0/V$ , and the inequality $t>>\tau _0$ is supposed.

From here it follows that the correction is forced to include additional
logarithmical dependence of $\Phi (0)$ and so of $G(q,t)$ on time. This
dependence can be separated as the multiplier
\[
A(t)\approx [\ln \frac t{\tau _0}]^{-1}
\]
if make the change $\Phi (\frac q{Vt})\Rightarrow A(t)\Phi (\frac q{Vt})$
accompanied by the requirement $\Phi (0)=1$ .

If all the cumulants and statistical moments of $\Delta Q$ are
finite then $\Phi (x)$ should decrease from unit to zero
in a sufficiently fast way.
Details of this function are of no principal importance. For example, one
can choose $\Phi (x)=\exp (-x)$ . Finally, the simple estimates of the
integral give

\[
\left\langle \exp \{ik\Delta Q\}\right\rangle \approx
\exp \{-ikU_0tA(t)\ln (\frac{\tau _0}t-ikQ_0)\}=
\]
\[
=\exp \{ikU_0t-ikU_0tA(t)\ln (1-ikVt)\}
\]
under the conditions $t>>e\tau _0$ and $\left| k\right| Q_0<<1$ , with the
latter corresponding to $\Delta Q>>Q_0$ .

Evidently, this is the CF of probability distribution which is as much close
to ideal Cauchy distribution as the limitations of statistical moments do
allow. There is also slightly other way to avoid the divergency [8], which
results in the CF $\left\langle \exp \{ik\Delta Q\}\right\rangle \approx
\exp \{ikU_0t[1-ikVt]^{-A(t)}\}$ . The corresponding probability
distribution resembles Levy-Khinchin stable distributions with Levy's
alpha-parameter close to 1, $\alpha _L=1-A(t)$ , and practically reduces to
the above Cauchy type distribution if $A(t)<<1$ . One more possible
modification was obtained in [10].

\subsection{LONG-RANGE STATISTICS OF PURELY\\ RANDOM FLOW OF EVENTS}

The time scale $\tau _0$ should be treated just as life-time of the memory
of random walk: at $t>>\tau _0$ the transition to almost scale invariant
behaviour takes place. At once, the factor $\frac{\tau _0}t$ in logarithm
serves as the cut parameter. If we neglected it the CF would transform into

\[
\left\langle \exp \{ik\Delta Q\}\right\rangle \approx \exp \{-ikU_0tA(t)\ln
(-ikQ_0)\}
\]
what corresponds to pure Cauchy statistics with infinite moments.

The probabilistic properties of $\Delta Q$ significantly depend on the
parameter $\kappa (t)\equiv V/[U_0A(t)]=Q_0/[\tau _0U_0A(t)]$ . If $\kappa
<<1$ then the distribution is almost Gaussian. In opposite case, $\kappa >>1$
, it approaches to ideal asymmetric Cauchy distribution and its density has
the long tail $\approx U_0tA(t)\Delta Q^{-2}$ although cut at far $\Delta Q$
values [6,8]. In this case all the probabilistic properties become almost
insensitive to how far the long tail is cut. For example, the estimates take
place $P\{U>U_0\}\approx \frac 1\pi \arctan (\pi /\ln \kappa )$ , and
$P\{U>2U_0\}\approx A(t)\exp (-U_0\tau _0/Q_0)$ where $P\{\}$ denotes the
probability.

{\it The essential feature of the statistics is a difference between the
mean and most probable values of }$\Delta Q${\it \ as well of the rate }
$U=\Delta Q/t${\it \ }. In particular, at $\kappa >1$ this difference is
decsribed by $U_0-U_{mp}\approx U_0A(t)\ln \kappa (t)$ , with $U_{mp}$ being
the most probable rate.

\subsection{CAUCHY STATISTICS AND 1/F-NOISE}

In contrary to probabilities, the second and higher statistical moments are
exteremely sensitive to the long tail being determined mainly by bad
untypical occurences. From the obtained CF one gets
\[
\left\langle \Delta Q^2\right\rangle -\left\langle \Delta Q\right\rangle
^2\approx 2U_0\frac{Q_0}{\tau _0}t^2A(t)\,\,,\,\left\langle U^2\right\rangle
-\left\langle U\right\rangle ^2\approx 2U_0\frac{Q_0}{\tau _0}A(t)\,\,
\,\,(\,t>>\tau _0)\]

Therefore, the variance of time-averaged rate $U(t)$ tends to zero but only
in logarithmically slow way. This means that the rate undergoes
low-frequency 1/f fluctuations. It is easy to show that their relative
spectrum is $S_{\delta U}(f)\approx (Q_0/U_0\tau _0)f^{-1}(\ln 2\pi f\tau
_0)^{-2}$ , at $f\tau _0<<1$ .

The only free dimensionless parameter of the statistics is $Q_0/U_0\tau _0$
. It determines both the 1/f-noise level and the degree of non-Gaussianity
of the statistics. Therefore, the close relations between probabilistic
properties and spectral properties of memoryless random flow of events do
exist, and 1/f-noise is as strong as non-Gaussian. Both the 1/f spectrum and
bad statistics mutually come from the absence of long-living memory. Their
quantitative connection is the manifestation of highly irregular fractal
structure of realizations of $\Delta Q(t)$ [8] similar to that of Levy
flights with $\alpha _L\approx 1$ .

\subsection{ABSOLUTE TIME SCALES AND SCALELESS FLUCTUATIONS}

In general, any random walk subjected to a stable statistics [12] possesses
self-similar fractal structure. But, unfortunately, a pure mathematical Levy
flight has no intrinsic time scale at all, and so its rate looks as a white
noise. Besides, at alpha-parameter $\alpha _L\leq 1$ even its mean value
turns into infinity. This is too idealized model to apply it to real
processes which always have some absolute lower time scale. We showed that
in the case $\alpha _L=1-0$ the existence of such a scale automatically
leads to appearance of logarithmically time-dependent factors like $A(t)$
responsible for long-living scaleless statistical correlations. One comes to
the idealized statistics only in the limit $\tau _0\rightarrow 0$ ,
$U_0\propto 1/A(t)\rightarrow \infty $ .

\subsection{LOW-FREQUENCY NOISE AS DETERMINED\\ BY SHORT-TIME SCALES}

Formally $Q_0/U_0\tau _0$ is arbitrary parameter, but in concrete
applications it is possible to suggest some natural estimates for it. For
example, let $\Delta Q(t)$ be formed by well distnguishable identical
discrete events (dozes, electron jumps, etc.). In this situation it is
natural to choose $Q_0=1$ , and $U_0\approx 1/\tau _c$ , with $\tau _c$
being the mean time separation of events, that is the correlation time of
the Poissonian short-range statistics of $u(t)$ . Then $S_{\delta
U}(f)\approx (\tau _c/\tau _0)f^{-1}(\ln 2\pi f\tau _0)^{-2}$ , and the
relative 1/f-noise level is determined by ratio of mean distance between
events and the memory duration, i.e. by the amount, $\sim \tau _0/\tau _c$ ,
of recent events which could be kept in memory.

The corresponding mean square relative uncertainty of the rate if measured
during $t>>\tau _0$ is $\left\langle \delta U^2\right\rangle \sim (\tau
_c/\tau _0)(\ln t/\tau _0)^{-1}$ . {\it Hence, the rate can not be predicted
with much better precision ( }$<<\sqrt{\tau _c/\tau _0}${\it \ ) than what
is available for the system itself and achievable by its self-observation
during its memory life-time.}

The particular case $\tau _0\approx \tau _c$ corresponds to minimally short
memory, with low-frequency noise $S_{\delta U}(f)\approx \frac \alpha f$ ,
where $\alpha =(\ln 2\pi f\tau _0)^{-2}$ , $\alpha \sim 0.02$ at $f\tau
_0\sim 10^{-4}$ and $\alpha \sim 0.002$ at $f\tau _0\sim 10^{-9}\div
10^{-13} $ (what is typical for electric noises). It can be shown
that at $\alpha >0.0001$ the probability density of $U$ has noticable
long tail.

The constructed model is able to give satisfactory quantitative estimates
[6-8] of both the spectrum of 1/f-noise and its statistics which can be
experimentally investigated [22-24] in small carbon resistors, thin metal
whiskers and some other systems (in particular, a shape of experimental
probabilistic histograms and the difference between average and most
probable values can be correctly described).

\subsection{1/F-NOISE IN DNA}

Now we are able to hypothetically interprete also the example ix) [17,18].
The 1/f-noise in a distribution of any given sort of bases over DNA sequence
could mean merely that only more or less localized pieces of DNA serve for
encoding an information, while the amounts of bases accumulated by long
pieces and total number of given bases in DNA are of no importance.

If this is true, we concern only statistical long correlations which
themselves do not have any specific significance. But, in accordance with
above phenomenology, the magnitude of 1/f-noise could contain an information
about typical length of actually self-connected pieces.

In any case, the difference between statistical and actual (causal,
structural) correlations should be taken into account. The paper [18] is one
a very few works we know where this difference is realized.

\subsection{COIN TOSSING AND LASER BEAM}

Of course, the statistics under consideration is only the principal
possibilty, not a necessity. Everybody knows about random flows of events
which keep no place for 1/f-noise. For instance, the mathematical coin
tossing first investigated by J.Bernoulli in his ''Ars Conjectandi'' in 1713.

However, one could say nothing about flipping a physical coin tossing, if do
not assume that the events are  ''statisticaly independent'' and the
probabilities of heads are one half, thus coming back to the mathematical
model. May be, the probabilities differ from one half, but this is of no
importance if they are thought as purely certain. The essence of the
''statistical independency'' is that one identifies the probabilities of
heads with the ''the probabilities per unit time'' (i.e. per unit tossing)
which eventually coinside with limits of averaging over many tossings.

Therefore, to say that events are statistically independent is the same as
to assume that these limits exist and besides are achieved in sufficiently
fast way (not in logarithmical one). Hence, the only factual Bernoulli's
achievement was the law what describes the approach to the limit, provided
it does exist, not the proof of its existence.

What is for the physical coin, may be, in this case the flicker floor is
really negligible. But the opposite reality does not imply that there are
some slow processes, it means rather that the ideal certainty of the
probabilities materialized in ideal constancy of the coin takes no place.

In fact, the ideal constancy of coin serves as infinitely long-living causal
correlation what just ensures the absence of 1/f-noise in the model. The
Poissonian statistics of photons in an ideal laser beam is the more modern
example of Bernoullian flow of events. The absence of 1/f-noise in photon
flow in ideal beam also is due to hidden infinitely long causal correlation,
namely, to what all photons belong to one and the same quantum state.
Similarly, an electric supercurrent has no 1/f fluctuations, in contrary to
normal current in superconductors.

\subsection{QUANTUM DECAY}

From the viewpoint of conventional statistical interpretation of quantum
mechanics, the random flow of decays mentioned in the example a) is very
similar to coin tossing. One needs in definite probability of decay per unit
time and, besides, in statistical independency of decays in order to predict
the Poissonian statistics of decays. The 1/f fluctuations (observed at
frequencies below $10^{-3}\div 10^{-4}\,Hz$ ) are in principal contradiction
to this theoretical scheme. So we can conclude that the hypothesis about the
statistical independency is wrong. In other words, the decay probability has
no absolutely definite value. The observed 1/f-noise is quite equivalent to
1/f fluctuations of probability of decay.

In fact, one lacks the statistical independency in order to connect the
decay probability calculated in quantum mechanics and imaginary ensemble of
identical experiments. But if we wanted to prove the statistical
independency of two events we would need in the more complicated assumption
that any two pairs of events are statistically independent, and so on up to
infinity. Therefore the statistical independency can not be grounded in a
logical way. This circumstance is noted in the book [25].

It is the open question whether an uncompleteness of quantum mechanics or
primitivity of decay models is responsible for the discrepancy. Another
problem was pointed out by Prigogine [26]. It is well known that time
dependence of the decay probability includes non-exponential (power-law)
tail. It relates to times much larger than the life-time and so it itself is
practically unobservable. Nevertheless, it is inconsistent with the
principle that quantum decay does not remember the past: if one knows that a
nucleus has not yet decayed then any expectations for future should not
depend on how long it was being safe. As a consequnce, the probability of no
decay should be purely exponential, $P(t)=\exp (-\gamma t)$ .

In our opinion, it may be possible to avoid at once both the contradictions
in the unified way if suppose (or better deduce from quantum mechanics) that
the probability of decay per unit time, $\gamma $ , is not purely certain,
and that its uncertainty is just as large as necessary in order to disguise
the nonexponential tail. In other words, we should make difference between
the most probable behaviour of experimentally registered flow of decays and
its ensemble-averaged behaviour, as in general in memoryless flows of
events. Then the quantitative measure of this difference is nothing but that
of 1/f-noise.

\subsection{EQUILIBRIUM THERMAL 1/F-NOISE}

The equilibrium voltage noise in a disconnected tunnel junction is
constituted by random jumps of electrons from left to right-hand side of the
junction and in opposite direction. Let $N_{+}(t)$ and $N_{-}(t)$ be the
numbers of these jumps during some time interval $t$ . If $N_{+}>N_{-}$ then
the electric voltage arises between sides, $u=e(N_{+}-N_{-})/C$ , with $C$
being their electric capacity, which makes one sort of jumps more probable
than another. Due to this thermodynamical opposite reaction, after a
characteristic relaxation time of order of $\tau _c=RC$ , with $R$ being the
junction resistance, the system returns to a vicinity of the chargeless
state. Therefore the difference $N_{+}-N_{-}$ always is kept near its
average zero value, and the time-smoothed current $e(N_{+}-N_{-})/t$ turns
to zero at $t>>\tau _c$ .

But let it happened that the sum $N(t)\equiv N_{+}+N_{-}\approx 2N_{+}$
essentially deviates from its average value, while the difference remains
close to zero. Such a fluctuation causes neither a charging nor any other
thermodynamical consequences. All stay as before! So we can expect that
later this fluctuation will not be ''compensated and damped'' by some
oppositely aimed process. Therefore, in reality the time-smoothed
value $N(t)/t$ has no reasons to tend to its ''theoretical''
ensemble averaged
value, moreover, it have no explicitly predictable limit at all [10].

Now notice that both types of jumps give an equal positive contribution to
the noise intensity. The latter is determined by the summary number of jumps
per unit time, $S=e^2N/t=e^2(N_{+}+N_{-})/t$ , with $S$ being the
instantaneous (time-averaged) power spectrum. Hence, we come to white noise
but whose intensity undergoes low-frequency 1/f-fluctuations, i.e. to
flicker "noise of noise".

\subsection{1/F-NOISE AS UNCERTAINTY\\ OF KINETICAL VALUES}

It is important to underline that in all the above examples the values
subjected to 1/f-fluctuations could be qualified as kinetical values. This
is principal peculiarity of 1/f-noise without slow processes. In contrary to
dynamical and thermodynamical variables, kinetical values characterize the
statistics of random transitions between various states of the system, not
statistics of the states themselves.

In general, good ergodicity with respect to visiting the states does not
imply the same with respect to transitions. Even in the simplest case, when
a model distinquishes only two states and each of them occupies well certain
portion of time, the rate of random jumps between the states may be
uncertain.

None kinetical value could be referred to a definite time moment. The matter
is that any kinetical value takes a shape only during a finite time interval
longer than memory life-times. This circumstance means that kinetical values
have no their own characteristic time scales. As a consequence, their
fluctuations naturally acquire a scale invariance with a scaleless power-law
low-frequency spectra.

\subsection{1/F-NOISE IN NON-EQUILIBRIUM STATE}

In thermodynamical systems every sort of kinetical events (random
transitions from one state to another) is accompanied by time-reversed
events, and a freedom of random behavior is restricted by definite ratio of
amounts of opposite events (by detailed balance, in equilibrium case). This
restriction arises from reversibility of underlying dynamics, always are
under thermodynamical control and can be taken into account with the help of
fluctuation-dissipation relations. The example is the
ratio $N_{-}/N_{+}\simeq 1 $ in equilibrium electric junction.

But fluctuations in the summary rate of direct and reversed kinetical events
does not destroy their balance and so does not cause complete counteraction.

If an external voltage $U$ is applied to the junction, then undumped
fluctuations in the total rate of jumps result in not only noise of noise,
but also strictly in the voltage and current. Now the relation $N_{-}\simeq
N_{+}\exp (-eU/T)$ is thermodynamically kept, instead of $N_{-}\simeq N_{+}$
, and the current, $e(N_{+}-N_{-})/t=e\tanh (eU/2T)N/t=\tanh
(eU/2T)S/e\propto S$ , begins to reflect uncertainty of the total number of
jumps per unit time. In this nonequilibrium state the voltage spectrum
acquires 1/f contribution in addition to the white background, and 1/f-noise
can be viewed directly in the two-point correlators as effective resistance
fluctuations.

\section{BROWNIAN MOTION REVISED}

\subsection{QUAZI-GAUSSIAN RANDOM WALKS}

Let $u(t)$ be equilibrium noise short-correlated in the sense of two-point
correlators. Then the integral $Q(t)=\int_0^tu(t)dt$ is symmetrical random
walk like Brownian motion. Its characteristic fractal properties are
determined by the scaling $Q(t)\propto \sqrt{t}$ , at least if $t$
sufficiently exceeds $\tau _0$, with $\tau _0$ being maximal correlation
time of $u(t)$. As it was shown in [4,5], if only one excludes very formal
case of ideally Gaussian $u(t)$ , then this scaling results in 1/f-noise of
instantaneous power spectrum, $S(t)$ , of the noise $u(t)$ and in definite
non-Gaussian long-range statistics of $Q(t)$ (termed quazi-Gaussian) which
resembles the statistics of Levy flights with alpha-paremeter slightly
smaller than 2, $\alpha _L=2-0$ . One of many examples is the equilibrium
voltage noise in disconnected electric junction. Another example is the
equilibrium Brownian motion, with $u(t)$, $Q(t)$ and $D(t)\equiv S(t)/2$
being the velocity, displacement of a Brownian particle and its current
diffusivity, respectively.

Principally, the statistics of $Q(t)$, like that of the above analysed
one-directed random walk, is derivable from rather trivial requirements to
be satisfied asymptotically, at $t>>\tau _0$.

a) The standard diffusion law takes place, $\left\langle Q^2(t)\right\rangle
=2D_0t$ .

b) $Q(t)$ behaves as infinitely divisible random value.

c) It approximately behaves as random walk with independent increments.

d) Due to forgetting the past the system does not distinguish what is a
proper rate of the walk and what is deviation from it. Any currently
happened location of the walk is equally permissible as starting point for
further adventures! Therefore, the scaling $Q(t)\propto \sqrt{t}$ is
peculiar for the statistics as a whole (not only for the mean square
displacement). This can be expressed in the form of approximate statistical
identity
\[
Q(\lambda t)\sim \sqrt{\lambda }Q(t)
\]

These requirements imply nearly stable probability distribution whose
characteristic function (CF) looks as [5,7,8]
\[
\left\langle \exp \{ikQ\}\right\rangle \approx \exp \{D_0k^2tA(t)\ln (
\frac{\tau _0}t+k^2q_0^2)\}=
\]
\[
=\exp \{-D_0k^2t+D_0k^2tA(t)\ln (1+D^{\prime }k^2t)\}
\]
Here $A(t)\approx 1/\ln (t/\tau _0)$ , $t>>\tau _0$ , $k^2q_0^2<<1$ ,
$D^{\prime }=q_0^2/\tau _0$ , and the scales $q_0$ and $\tau _0$ serve as
lower spatial and temporal bounds of the scale invariance of the random walk.
      Under the formal limit $\tau_0\rightarrow 0$ , $D_0\propto
1/A(t)\rightarrow \infty $ , this CF turns into CF of pure quazi-Gaussian
distribution, $\exp \{-Dtk^2\ln (1/k^2q_0^2)\}$ , with infinite second and
higher even-order moments and without characteristic time at all. This
random walk constructed in [4,5] is absent in standard classification of
Levy flights [12]. Independently on us it was found in the mathematical work
[27]. But many of applications need in random walk models which possess
finite statistical moments and non-zero absolute lower time scale with
consequently deformed scale invariance, as in the model under consideration.

\subsection{FOUR-POINT CUMULANTS\\ AND DIFFUSIVITY FLUCTUATIONS}

The main information about 1/f-noise is contained in unusually behaving
4-order cumulant of the displacement what follows from the above CF,
\[
\left\langle Q^{(4)}(t)\right\rangle \equiv \left\langle
Q(t),Q(t),Q(t),Q(t)\right\rangle \propto t^2A(t)\approx t^2[\ln (t/\tau
_0)]^{-1}
\]

In general, the standard variance $\left\langle Q^2(t)\right\rangle =2D_0t$
can be accompanied by non-standard 4-point cumulant like $\left\langle
Q^{(4)}(t)\right\rangle \propto t^{1+\gamma }$ with $\gamma >0$.
At any $\gamma >0$ such the asymptotics is formally equivalent
to low-frequency
fluctuations of instantaneous power spectrum (diffusivity) of $u(t)$ .
Indeed, by the formal definition of cumulants,
\[
\left\langle \exp [ikQ(t)]\right\rangle =\exp [-k^2\left\langle
Q^2(t)\right\rangle /2+k^4\left\langle Q^{(4)}(t)\right\rangle /24-...]
\]
From the other hand, one can phenomenologically treat $u(t)$ as a shot noise
what would be Gaussian if its power spectrum was constant, $S(t)=2D_0$ , but
becomes non-Gaussian because of its relatively slow random modulations,
$S(t)=2D(t)$. Correspondingly, the CF can be represented by double averaging,
first over fast Gaussian noise under fixed realization $D(t)$ and then over
random $D(t)$ :
\[
\left\langle \exp [ikQ(t)]\right\rangle \approx \left\langle \left\langle
\exp [-k^2\int_0^tD(t^{\prime })dt^{\prime }]\right\rangle \right\rangle =
\]
\[
=\exp [-k^2D_0t+k^4\int_0^t\int_0^tK_D(t_1-t_2)dt_1dt_2/2-...]
\]
Here the doubled brackets denote the second stage of averaging, $D_0\equiv
\left\langle D(t)\right\rangle $ , the inequality $t>>\tau _c$ is implied,
and $K_D(\tau )$ is the effective correlation function of fluctuating
diffusivity $D(t)$ ,
\[
K_D(t)=\frac 1{24}\frac{d^2}{dt^2}\left\langle Q^{(4)}(t)\right\rangle
=\frac 12\int_0^t\int_0^t\left\langle u(t),u(\tau ^{\prime }),u(\tau
^{\prime \prime }),u(0)\right\rangle d\tau ^{\prime }d\tau ^{\prime \prime }
\]

The unusual 4-point cumulant means slow asymptotics $K_D(\tau )\propto \tau
^{\gamma -1\text{ }}$ and thus $f^{-\gamma }$ low-frequency spectrum of
diffusivity fluctuations. The case $\gamma >1$ formally corresponds to
non-integrable spectrum like that of nonstationary random process, but this
is only imaginary non-stationarity because $u(t)$ is purely stationary
random process (all its cumulants possesses time translation invariance).

\subsection{MICROSCOPIC TIME SCALES\\ AND MAGNITUDE OF 1/F-NOISE}

In case of quazi-Gaussian random walk $\gamma \rightarrow 1$ , $K_D(\tau
)\propto A(\tau )$ , and the low-frequency part of spectrum of relative
diffusivity fluctuations is $S_{\delta D}(f)\approx \alpha (f)/f$ ,
where $\alpha (f)\equiv (q_0^2/D_0\tau _0)[\ln 2\pi f\tau _0]^{-2}$ .

Let us estimate this noise taking in mind literally Brownian motion of a
microscopic particle. The mean diffusivity can be expressed as $D_0\approx
q_c^2/\tau _c$ with $q_c$ and $\tau _c$ being its mean free path and free
flight time, respectively. Clearly, $\tau _c$ coinsides with correlation
time of the velocity $u(t)$. It is quite reasonable to suppose that spatial
scale invariance takes beginning just from one free path,
that is $q_0\approx q_c$ . Then magnitude of 1/f-noise is determined by
the ratio $\tau _c/\tau _0$.

Again we see that 1/f-noise is as strong as long is the memory duration as
compared characteristic time separation of elementary kinetical events
(interactions, collisions, etc.). The smaller is a number of previous
kinetical events remembered and taken under control by a mechanism what
produces white noise the greater is the accompanying 1/f-noise. In fact, one
can be convinced of validity of this statement [7] with respect to different
particular cases of Brownian motion of charge carriers in semiconductors and
metals, to charge transport in normal junctions and superconducting
Josephson junctions, and to Brownian particle in liquids (in the latter
cases the difference between $q_0$ and $q_c$ can be significant too).

The special case, $\tau _0\approx \tau _c$ , when scale invariance starts
immediately after one free flight time, corresponds to natural basic level
of diffusivity noise. This is the case for electrons in a slightly doped
semiconductor, due to inelasticity of collisions. Taking into account
that $\tau _c\sim 10^{-12}\,s$ , at $f\sim 1\,Hz$ (as well as at
frequencies a few
orders larger or smaller) one finds $S_{\delta D}(f)\approx 0.0015/f$ .
This estimate [4,5] is in good agreement with empirical Hooge
constant $\alpha \approx 0.002$ [1,2].

\subsection{PROBABILITY DISTRIBUTION\\ OF BROWNIAN MOTION}

In contrary to 4-th and higher cumulants, the probability density $W(Q,t)$
of quazi-Gaussian random walk always has only small quantitative differences
from that of ideal Brownian motion, $W_G(Q,t)=(4\pi Dt)^{-1/2}\exp
(-Q^2/4Dt) $. If $D^{\prime }>D_0$ then $W(Q,t)$ has noticable power-law
tails whose extent approximately equals to $\sqrt{2D^{\prime }t}$ but
magnitude does not depend on $D^{\prime }/D_0$ , i.e. on the 1/f-noise
level. In the limit $D^{\prime }/D_0\rightarrow \infty $, for the
region $Q^2>>\left\langle Q^2\right\rangle $, one can get
$W(Q,t)\approx W_G(Q,t)+2D_0tA(t)/\left| Q\right| ^3$ .
From here the probability $P$ of
that $Q^2$ is three times larger than $\left\langle Q^2\right\rangle =2D_0t$
can be estimated as $P<P_G+[3\ln (t/\tau _0)]^{-1}$ , where $P_G$ means
similar probability for ideal Brownian walk, $P_G\approx 0.1$ . Therefore,
at $t/\tau _0>10^3$ the difference between $P$ and $P_G$ never
could exceed $0.05$. Analogouosly, one can find that at $t/\tau _0>10^3$
always $P\{\left|
Q\right| >4\sqrt{2D_0t}\}\approx [16\ln t/\tau _0]^{-1}<0.01$ .

\subsection{STATISTICAL SENSE OF LONG\\ CORRELATIONS OF KINETICAL
VALUES}

Hence, the probability of noticably far deviations from standard behaviour
does not exceed one percent even in the worst case (infinite 1/f-noise).
Nevertheless, just such rather unprobable occasions determine the
asymptotics of $K_D(t)$ and the level of diffusivity 1/f-noise.

Evidenly, the long tail $K_D(t)\propto A(t)$ reflects only the statistical
weight of untypical realizations of the random walk, not some actual
correlations inside realizations. More explicitly, any concrete realization
looks as sometimes typical and sometimes rare, and because of its fractal
structure every variant of its behaviour during longer time interval is
constituted by both typical and rare contributions from shorter subintervals.

The same can be said also about fluctuations of other kinetical values. From
the viewpoint of underlying dynamics, their correlation functions are
roughly phenomenological characteristics. However, one can present rigorous
definitions for these correlators in terms of four-point (or higher order)
cumulants (or related ensemble averages) of purely dynamical variables.

\section{DYNAMICS AND 1/F-NOISE}

\subsection{SOME QUESTIONS}

Of course, we are highly interested in 1/f-noise derived from Hamiltonian
dynamics. Unfortunately, none of realistic Hamiltonian models allows for
explicite calculations of 4-point cumulants even for Hibbsian equilibrium
ensemble. Moreover, this problem never was under careful consideration.
Instead, everybody wanted to somehow reduce the statistics to two-point
correlators thus diminishing specific contribution of 4-point cumulants.
Various means were attracted to this aim, either ansatzes like molecular
chaos, random phases, thermodynamical limit, continuous spectra of energy
levels, diagonal singularity and others, or exclusion of some interactions
what do not influence low-order correlations under interest.

Some of well investigated low-dimensional nonlinear mappings have the
intermittency regimes accompanied by 1/f-noise, but this noise is connected
with long periods of regular motion, that is with long memory and slow
processes (for example, see [28]). Therefore it is not surprising that this
1/f-noise becomes destroyed (saturated at zero frequency) under influence of
external white noise. So it is not the 1/f-noise we are interested in whose
low-frequency behaviour should not be supressed by noisy perturbation.

At present the axiomatic theory of the so-called Anosov systems is developed
[29,30] based on works by Sinay on scattering billiards and ideas by Anosov.
These systems are characterized by exponential instability and divergency of
phase trajectories and their mixing in phase space, possibly in company with
time reversibility and phase volume conservation (what is most interesting
for us). In such systems 1/f-noise could be connected with random temporal
non-uniformity of the rate of exponential instability. As far as we know,
this hypothetical possibility was not under investigation. In contrary, the
attention was concentrated on the systems and phase space divisions which
produce Bernoullian flows of events.

However, as it is was shown by Ornstein [13], in principle generalized
random flows of events (K-flows) do exist which can not be recoded into
Bernoullian flows. One could search for 1/f-noise in concrete realizations
of such generalized flows, but this subject is still apart of achievements
of the theory.

\subsection{WHY KINETICS LOST 1/F-NOISE ?}

Because none realistic dynamics can be exactly solved, physicists resort to
kinetical models of relaxation, noise and irreversible phenomena. Everybody
knows how this is carried out. First one chooses a reasonable coarsened
characterization of the system, secondly, analyses what variants of its
evolution may occur during some short time step $\Delta t$ if start from an
uncompletely described state, then evaluates probabilities of the variants
and finally keeps these probabilities as governing parameters for a
long-time evolution during many steps.

This theoretical scheme was expressively criticized by M.Kac [31]. Clearly,
its main defect is its last ansatz, because it means that probability of any
chain of transitions between states is equalized to product of probabilities
of marginal transitions, i.e. that successive transitions are considered as
statistically independent events. One could say that such a model divides
the real trajectories to short pieces and randomly permutes these pieces at
each time step $\Delta t$. As a consequence, the model can correctly
describe short correlations and relaxation, but in general is unable to
correctly reproduce a long time behavior of trajectories.

Usually one sees nothing criminal in this simplification. But we insist that
long-living statistical dependencies and non-Bernoullian behaviour may mean
merely that trajectories take liberties just because of no really
long-living memory. Then the statistical independency looks as too rough
approximation. Eventually, this approximation results in the loss of
1/f-noise.

We see that modern theory is not kind with respect to 1/f-noise.
Nevertheless, we have pointed out two Hamiltonian systems, both being basic
models of statistical mechanics, which produce 1/f-fluctuations of kinetical
values, namely, a slightly non-ideal gas [9] and slightly unharmonic
phononic system [11]. In our opinion, every many-particle Hamiltonian model
which really bears a relaxation inevitably bears also 1/f fluctuations of
the rate of relaxation.

\subsection{EQUILIBRIUM $1/F^2$-NOISE IN\\ THE KAC'S RING MODEL}

Many years ago M.Kac [31] suggested the model which represents a good toy
analogy of the problems always accompanying derivation of relaxation, noise
and kinetics from dynamics. The system is the ring chain of $n>>1$ boxes
each containing the only black or white ball. At every moment of discrete
time $t=..-1,0,1,2...$the balls displace to left along the ring. The boxes
are characterized by time-independent parameters $\sigma _j=\pm 1$ .
If $\sigma _j=-1$ then $j$-th box inverts the colour of any ball what
leaves it.
The dynamical variables $\xi _j(t)=\pm 1$ are defined so that $\xi _j(t)=1$
if $j$-th box contains a white ball at time moment $t$ and $\xi _j(t)=-1$ if
it contains a black ball. The equations of motion are easily soluble:
$\xi_j(t+1)=\sigma _{j-1}\xi _{j-1}(t)$ (the number $j=-1$ is
equivalent to $j=n-1$). The equilibrium statistical ensemble is defined
by conditions that $\xi _j$ are statistically independent if considered
at one and the same time
moment and $\xi _j=\pm 1$ with equal probabilities $1/2$ , and that constant
parameters $\sigma _j$ equals to $1$ or $-1$ with probabilities $p,1-p$ also
being independent one on another. Evidently, the dynamics is time-reversible
and the ensemble is time-invariant. We confine ourselves by the simplest
case $p=1/2$ .

The quantity under macroscopic observation is the mixed colour $u(t)\equiv
[\sum_j\xi _j(t)]/\sqrt{n}$ . It is easy to verify that $\left\langle
u(t)\right\rangle =0$ and that from the point of view of second-order
correlators $u(t)$ represents a white noise: $\left\langle
u(t_1)u(t_2)\right\rangle =\delta _{t_1t_2}$ , with Kronecker symbol on the
right-hand side.

Kac demonstrated [31] that this model allows for wide analogies with serious
statistical mechanics and for illustrations of approximate methods usually
attracted when building kinetics. But Kac did not note that $u(t)$ is very
far from white noise in the sense of higher-order statistics.

Let us consider the discrete random walk $Q(t)=\sum_{k=t_0+1}^{t_0+t}u(k)$
at $1\leq t<n$. Obviously, $\left\langle Q^2(t)\right\rangle =2D_0t$, with
diffusivity $D_0=1/2$. If this process was asymptotically gaussian and
subjected to the central limit theorem at $t>>1$, the four-order cumulant of
$Q(t)$ would grow also as a linear function of time, $\left\langle
Q^{(4)}(t)\right\rangle \propto t$. For instance, if $u(t)$ was purely white
noise, one would get $\left\langle Q^{(4)}(t)\right\rangle =-2t/n$ .
However, the explicit calculation lead to
\[
\left\langle Q^{(4)}(t)\right\rangle \equiv \left\langle Q^4(t)\right\rangle
-3\left\langle Q^2(t)\right\rangle ^2=2(2t^2-1)(t-2)/n\propto t^3\,\,(t>>1)
\]
(this easy reproducable calculation was performed ten years ago and sent to
1988 Conference on noise in physical systems in Vilnius, but, unfortunately,
was not accepted either there or for journal publications).

This result proves the existence of infinitely long-living four-point
correlations (cumulants) of the noise $u(t)$, and means that standard
kinetical approaches can not present a qualitatively correct approximation
(let being quantitatively good) of the noise. From the viewpoint of external
observer, such 4-th cumulant looks as manifestation of low-frequency
fluctuations of diffusivity, i.e. of instantaneous power spectrum of $u(t)$.
In accordance with the above formulas, the corresponding spectrum is
$\propto f^{-2}$. The factor $1/n$ is analogous to what takes place in the
spectrum of relative resistance 1/f-noise in bulk conductors (example i) ).

\subsection{NON-STATIONARY NOISE\\ IN STATIONARY SYSTEM}

Perhaply, this noise could not be qualified as what is due to forgetting the
past, sooner the actual correlations carried in by random parameters $\sigma
_j$ do work. Nevertheless, the obtained $1/f^2$-noise of noise is
principally important by two reasons. First, it gives the example of
completely calculatable long-living higher-order correlations in
time-reversible dynamics. Secondly, it helps to understand why seemingly
non-stationary fluctuations of kinetical values can occur even in
equilibrium system.

With no doubts, the non-integrable spectrum $1/f^2$ belongs to stationary
system . Indeed, the statistics of $Q(t)$ does not depend on the arbitrary
initial time moment $t_0$. The spectrum $1/f^2$ reflects only the role of
duration $t$ in the measurement of kinetical value (diffusivity) and means
that the longer is the measurement the worse is its result. The similar
strange property characterizes Levy flights with $\alpha _L<1$ [12].

Of course, the guessing is that any kinetical value is delocalised in time.
Therefore its bad behaviour does not contradict to ergodicity of visits of
states. Moreover, we can expect that at fixed duration $t$ the four-point
cumulant calculated with the help of ensemble averaging coinsides with that
obtained from the time smoothing over $t_0$.

Besides, one should remember that non-stationarity in the sense of
statistical moments coexists with stationarity of the most probable
behavior. But because of non-stationarity the result of experiments can be
sensitive to details of the measuring procedure [8,9].

\subsection{1/F-NOISE OF FRICTION AND OF LIGHT\\ SCATTERING IN PHONONIC
SYSTEM}

A quartz crystal is the system where thermodynamically equlibrium and
nonequilibrium manifestations of one and the same 1/f-noise are observed by
different methods. As it is known for a long time [14,15], when quartz is
used as frequency stabilizator in electronic devices, its quality is
restricted by 1/f-fluctuations of inner mechanical friction and consequently
of dissipation. Not long ago the 1/f-fluctuations of intensity of the laser
light scattered by equilibrium quartz were discovered [16].

In [11] we showed that the statistical correlators which describe the
fluctuations of dissipation can be generally connected with four-point
equilibrium cumulants which describe thermal fluctuations of quantum
probabilities of photon scattering by phonons. It was shown also that both
the friction and light scattering 1/f-fluctuations can be reduced to thats
of the relaxation rates (life-times) of phononic modes. In their turn, this
fluctuations originate from the same phonon interactions what produce
exponential instability, continuous spectra and the relaxation phenomenon
itself.

The logical way to 1/f-noise looks as follows. Due to exponential
instability, the interaction of given test phononic mode with the thermal
bath consisting of other modes produces both its noisy disturbation and
damping. In zero approximation, the friction is referred to linear response
of the bath to test mode and is non-random, and correspondingly the noise is
Gaussian. The linear response seems good approximation because the influence
of the only test mode is $1/\sqrt{N}$ times smaller than summary
interactions inside the bath, with $N$ being total number of modes, $N\sim
10^{20}$ .

Next, the equations which govern the linear response look as equations of
coupled harmonic oscillators without friction but with fluctuating
couplings. Even being Gaussian this fluctuations cause a stochastic
instability of the response, in the sense of second and higher statistical
momemts. Therefore in better approximation the friction becomes random with
infinite variance and besides with infinitely long non-locality. As a
consequence, the infinitely long-living four-point statistical correlations
of phononic variables do occur.

In a more developed picture, one should consider the completely nonlinear
response. But, as it was argued in [11], the long-living many-point
correlations survive and correspond to giant low-frequency fluctuations of
phase diffusivities of phonon modes and thus of their relaxation rates
(relaxation times).

\section{GENERALIZED FLUCTUATION-\\-DISSIPATION RELATIONS}

\subsection{PROBABILISTIC RELATIONS}

In any theory concerning 1/f-noise in thermodynamical systems the exact
generalized fluctuation-dissipation relations (FDR) can be useful which were
considered in [32-34] (see also [7,9,11,38]). Except classical
fluctuation-dissipation theorem, Kubo formulas and Onsager relations, FDR
include infinitely many additional connections between nonlinear responces
and higher-order equilibrium and nonequilibrium cumulants.

All the variety of FDR can be expressed with only relation what follows from
the phase volume conservation and time reversibility of Hamiltonian
dynamics. For instance, in case of initially equilibrium Hibbsian canonical
ensemble later perturbed by some time-dependent variation of the
Hamiltonian, the producing FDR can be written [32-34] as
\[
\Pi \{Trajectory;Forces\}\exp (-Work/T)=\Pi \{\widetilde{Trajectory};%
\widetilde{Forces}\}
\]
There $\Pi \{Trajectory; $$Forces\}$
is the probability of realization of some
process ($Trajectory$) under given external
$Forces$=$Perturbations$+$Conditions$ (in the sense of probability
density in functional space), $Work$
is the work produced by the perturbation during this process
depending on $Trajectory$ and $Forces$ , $T$ is the temperature of
initial Hibbsian probability distribution, and
$\Pi \{\widetilde{Trajectory}; $$\widetilde{Forces}\}$
is the probability of realization of the time-reversed process under
time-inverted perturbations and conditions. The operation $\widetilde{}$
means the inversion of time direction and besides of signs of variables
which have the odd time parity. From here, after integration over all
possible processes, the remarkable equality $\left\langle \exp
(-Work/T)\right\rangle =1$ does follow what leads to inequality
$\left\langle Work\right\rangle \geq 0$. Particularly, if
the perturbation is described by Hamiltonian
\[
H(q,p,t)=H_0(q,p)-\sum Q_j(q,p)x_j(t) ,
\]
with $q$,
$p$ and $x(t)$ being canonical coordinates and momentums and external
forces, respectively, one has $Work=\int \sum J_j(t)x_j(t)dt$ ,
where $J_j(t)=\frac d{dt}Q_j(q(t),p(t))$ are the flows
(velocities, currents, etc.) conjugated with the forces.

This FDR is valid also in quantum case, under certain definition of quantum
characteristic and probability functionals [34]. Besides, this FDR can be
extended to nonequlibrium initial ensembles [34,38], in other words, to a
combination of dynamical perturbations (i.e. thats of Hamiltonian) and
thermal perturbations (i.e. thats of probability distribution in phase
space). The producung FDR looks quite similar besides that the factor $\exp
(-Work/T)$ should be replaced with $\exp (-\Delta S)$ where $\Delta
S\{t,Trajectory;Forces\}$ has the sense of increment of entropy (or similar
thermodynamic potential) during $Trajectory$.

\subsection{FLUCTUATIONS OF DISSIPATION}

Let $\left\langle J(t_1),...,J(t_n)\right\rangle _q$ denotes the $n$-th
order nonequilibrium cumulant corresponding to the time-cut modification of
perturbing forces, $x(t)\Rightarrow x(t)\eta (t-\min (t_1,...,t_n))$ , where
$\eta (t)=1$ at $t>0$ and $\eta (t)=0$ at $t\leq 0$ (thus the most early
variable belongs to yet equilibrium state). Such correlators were termed in
[34] quazi-equilibrium. Then the rigorous relation takes place [33,34,11],
\[
\left\langle J_j(t)\right\rangle =\frac 1T\int_{-\infty }^t\left\langle
J_j(t),J_k(t^{\prime })\right\rangle _qx_k(t^{\prime })dt^{\prime }
\]
what extends Kubo formulas to arbitrary non-linear responce. As a
consequence, the average value of the work (i.e. of the energy absorbed by
the system), $\left\langle Work\right\rangle =\int_{-\infty }^t\left\langle
J(t)\right\rangle x(t)dt$ , can be simply expressed via two-point
quazi-equilibrium correlators.

The similar formulas for higher cumulants can be obtained [34] which lead to
the formally exact expression for variance of the work [11] :
\[
\left\langle Work^{(2)}\right\rangle =2T\left\langle Work\right\rangle
+\frac 2T\int_{-\infty }^tx(1)x(2)\left\langle J(1),J(2),J(3)\right\rangle
_qx(3)d1d2d3
\]
where $\left\langle Work^{(2)}\right\rangle =\left\langle
Work^2\right\rangle -\left\langle Work\right\rangle ^2$, and for brevity the
time variables (and indices) are denoted by integers and triple time
integral is taken under condition $1>2>3$ . Here the
first term on the right-hand side describes shot noise accompanying the
average energy influx, and the second term involves an excess noise. The
latter enters by means of three-point cumulants, so one could not snatch at
it in Gaussian approximation. Under infinitesimally small perturbation,
$x\rightarrow 0$, the excess contribution is of order of $x^4$ and can be
expressed as

\[
\left\langle Work^{(2)}\right\rangle =2T\left\langle Work\right\rangle +
\]
\[
+\frac 2T\int_{-\infty }^tx(1)x(2)\left\langle J(1),\Gamma
(2,3),J(4)\right\rangle _0x(3)x(4)d1d2d3d4+O(x^6)
\]
where the dynamic linear differential responce function is introduced,

\[
\Gamma _{jk}(t,t^{\prime })\equiv \left[ \delta J_j(t)/\delta x_k(t^{\prime
})\right] _{x=0}
\]
the subscript 0 denotes averaging over equilibrium Hibbsian ensemble, and
integrations are ordered by the conditions
$1>4$, $2>3$, $2>4$, $3>4$.

If low-frequency fluctuations of kinetical values responsible for energy
dissipation take place then these fluctuations are reflected just the excess
noise term. In a steady state, when $\left\langle Work\right\rangle \propto
(t-t_0)x^2$ , with $t_0$ being the start time of perturbation, a $1/f^\gamma
$-noise should result in $\propto (t-t_0)^{1+\gamma }x^4$ behaviour of the
excess part of energy variance. Hence, nearly equilibrium fluctuations of
dissipative values (resistance, friction, mobility, etc.) can be generally
reduced to four-point equilibrium cumulants like $\left\langle J(1),\Gamma
(2,3),J(4)\right\rangle _0$ which describe specifically non-Gaussian
correlations between random responce and noisy flows. In their turn, these
correlators can be connected with literally four-point cumulants
$\left\langle J(1),J(2),J(3),J(4)\right\rangle _0$ , due to the four-point
FDR [32,11], in part due to

\[
\frac 1T\left\langle J(1),J(2),J(3),J(4)\right\rangle _0=\left[ \frac \delta
{\delta x(4)}\left\langle J(1),J(2),J(3)\right\rangle \right] _{x=0}=
\]
\[
=\left\langle J(1),\Gamma (2,4),J(3)\right\rangle _0+\left\langle \Gamma
(1,4),J(2),J(3)\right\rangle _0+\left\langle J(1),J(2),\Gamma
(3,4)\right\rangle _0
\]
where the time moments are ordered as $1>2>3>4$. More rich information is
given by the 7 independent four-point relations derived in [32], in
particular, by the FDR

\[
\frac 1{T^2}\left\langle J(1),J(2),J(3),J(4)\right\rangle _0=\left[ \delta
^2\left\langle J(1),J(2)\right\rangle /\delta x(3)\delta x(4)\right] _{x=0}+
\]

\[
+\left[ \delta ^2\left\langle J\widetilde{(2)},\widetilde{J(4)}\right\rangle
/\delta \widetilde{x(3)}\delta \widetilde{x(1)}\right] _{x=0}
\]
where the same time ordering is supposed, $1>2>3>4$,
and tilda means $\widetilde{x_j}(t_j)\equiv \varepsilon
_jx_j(-t_j)$ , $\widetilde{J_j}(t_j)\equiv -\varepsilon _jJ_j(-t_j)$ , with
$\varepsilon _j$ being the time parities of forces.

\subsection{CONNECTIONS BETWEEN EQUILIBRIUM\\ AND NON-EQUILIBRIUM RANDOM
WALKS}

Let $Q(t)=Q(q(t),p(t))$ be coordinate of some random walk and $x(t)$ be
generalized force conjugated with $Q$ in the above defined sense. In two of
many possible examples $Q(t)$ is a coordinate of Brownian particle with
$x(t) $ being the external force acting on it, or $Q(t)$ is electric charge
transported through some conductor with $x(t)$ being applied voltage.
Let $Q(t)$ be time-even variable, the force remains constant after its
switching
on ( $x=0$ at $t<0$ , $x=const\neq 0$ at $t>0$) and $Q=0$ at $t=0$ . Then
the general producing FDR results in simple relation [7,9,38],

\[
W(Q,t)\exp (-\frac{Qx}T)=W(-Q,t)
\]

\[
W(Q,t)-W(-Q,t)=\left[ W(Q,t)+W(-Q,t)\right] \tanh (\frac{Qx}{2T})
\]
where $W(Q,t)$ is the probability density of $Q$ at $t>0$. From here one
gets $\left\langle Q(t)\right\rangle =\left\langle Q(t)\tanh
[Q(t)x/2T]\right\rangle $ . The expansion of the latter relation over $x$ up
to $x^3$ yields $\left[ \frac \partial {\partial x}\left\langle
Q(t)\right\rangle \right] _{x=0}=\left\langle Q^2(t)\right\rangle _0/2T$ ,
and [9]
\[
\left[ \frac{\partial ^3}{\partial x^3}\left\langle Q(t)\right\rangle
\right] _{x=0}=\frac 3{2T}\left[ \frac{\partial ^2}{\partial x^2}%
\left\langle Q(t),Q(t)\right\rangle \right] _{x=0}-\frac 1{4T^3}\left\langle
Q(t),Q(t),Q(t),Q(t)\right\rangle _0
\]
with $T$ being the initial temperature [32].

The first of these formulas is nothing but Einstein relation between
low-field mobility and equilibrium diffusivity, or Nyquist relation, etc.
The second formula asserts the connection between quadratic non-equilibrium
addings to variance of $Q(t)$ (that is the excess noise), equilibrium fourth
cumulant and cubic response. The latter could be due to either nonlinear
dissipation mechanisms or to dependence on actual current temperature
maintained by Joulean heating. But, in any case, the left-hand side can not
grow in a more fast way than $\propto t$. On the contrary, in presence of
$f^{-\gamma }$ excess noise the first term on right-hand side should grow
as $\propto t^{1+\gamma }$. Hence, at sufficiently long time intervals
it must
be compensated by the second term, and we come to relations between
low-field excess nonequilibrium noise and four-point cumulant of equilibrium
noise. The result can be written also as $K_m(\tau )=\frac 1{T^2}K_D(\tau )$
[4,6,9], where $K_D(\tau )$ is the above defined correlator of diffusivity
and $K_m(\tau )$ is the mobility correlation function as introducedd by the
expansion $\left\langle Q(t),Q(t)\right\rangle =\left\langle
Q(t),Q(t)\right\rangle _0+x^2\int_0^t\int_0^tK_m(t^{\prime }-t^{\prime
\prime })dt^{\prime }dt^{\prime \prime }+O(x^4)$. This is the extension of
Einstein relation and similar relations to 1/f-fluctuations.

\subsection{UNIFICATION OF EQUILIBRIUM AND\\ NONEQUILIBRIUM BROWNIAN
MOTION}

The producing FDR can be reformulated in terms of characteristic function
(CF):
\[
\left\langle \exp [(ik-\frac xT)Q(t)]\right\rangle =\left\langle \exp
[-ikQ(t)]\right\rangle
\]
This is functional equation with respect to dependencies on $ik$ and $x$ .
It helps to find the connections between the whole statistics of
nonequilibrium random walk and that of equilibrium one, in presence of
1/f-noise, if we know something about statistics under neglecting 1/f-noise
[4,6,8,38].

One of possible solutions on the functional equation is $\left\langle \exp
[ikQ(t)]\right\rangle =\Xi (ikx/T-k^2,t)$ , with only two independent
arguments. It is easy to see that both the CF of quazi-Gaussian Brownian
motion and CF of asymmetric one-directed random walk with Cauchy statistics
can be unified into just such form [6,8],
\[
\left\langle \exp \{ik\Delta Q\}\right\rangle \approx \Xi
(ikx/T-k^2,t)\,\,,\,\,\Xi (\xi ,t)\equiv \exp \{-D_0\xi tA(t)\ln (\frac{\tau
_0}t-\xi q_0^2)\}
\]
Here $x=0$ corresponds to quazi-Gaussian case, while the change
$ikx/T-k^2\Rightarrow ikx/T$ , under the condition
$\left| k\right| x/T>>k^2$, leads to the second case. The condition
means that only the drift
component of the walk is observed which dominates at large time scales.
Comparing the CFs we find that the unification needs in relations
$U_0=D_0x/T $ and $Q_0=q_0^2x/T$ (or $V=D^{\prime }x/T$ ). These is nothing
but Einstein formula (or Nyquist formula, etc.) and its generalization for
1/f-fluctuations.

Besides, we obtain the statistical model of nonequilibrium Brownian motion,
with taking into account both the drift and diffusional displacements and
common 1/f-fluctuations of diffusuvity and mobility (in [6] FDR were used
even in order to derive the non-equilibrium Cauchy statistics from
quazi-Gaussian statistics obtained in [4,5]). The additional generalization
of the model allows for arbitrary dependencies $D_0=D_0(x)$ and $D^{\prime
}=D^{\prime }(x)$.

\section{CURRENT STATE AND PARADOXES\\ OF GAS KINETICS}

\subsection{BBGKY HIERARCHY}

The probabilistic formulation of the Hamiltonian dynamics of classical gas
is given by Bogolyubov-Born-Green-Kirkwood-Yvon (BBGKY) equations
\[
\left[ \frac \partial {\partial t}+v_1\frac \partial {\partial r_1}\right]%
F_1=\rho
\int
\Lambda_{12}
F_2dp_2dr_2
\equiv I_{12}(F_2)
\]
\[
\left[ \frac \partial {\partial t}+v_1\frac \partial {\partial r_1}+v_2\frac
\partial {\partial r_2}+\Lambda _{21}\right] F_2=\rho \int (\Lambda
_{13}+\Lambda _{23})F_3dp_3dr_3\equiv I_{13}(F_3)+I_{23}(F_3)
\]
\[
\left[ \frac \partial {\partial t}+L^{(n)}\right] F_n=\rho \int
\sum_j\Lambda _{jn+1}F_{n+1}dp_{n+1}dr_{n+1}\equiv \sum_jI_{jn+1}(F_{n+1})
\]
\[
L^{(n)}\equiv \sum_jv_j\frac \partial {\partial r_j}+\sum_{j>k}\Lambda
_{jk}\,\,,\,\,\Lambda _{jk}\equiv f(r_j-r_k)(\frac \partial {\partial
p_j}-\frac \partial {\partial p_k})
\]
Here $L^{(n)}$ is the $n$-particle Liouville operator,
$F_n=F_n(t,r_1..r_n,p_1..p_n)$ is the $n$-particle probability distribution
function (DF), normalized to whole (formally infinite) gas volume $\Omega $,
$\int F_ndp_ndr_n/\Omega =F_{n-1}$, the operators $\Lambda _{jk}$ decribe
pair interactions, with $f(r)$ being the interaction force, and $v_j=p_j/m$
with $m$ being the mass of particles.

\subsection{BOLTZMANN EQUATION}

More than 100 years ago Boltzmann suggested the phenomenological kinetical
equation to describe the evolution of rarefied gas with a short-range
repulsive interactions between particles. The Boltzmann equation (BE) can be
written as
\[
\left[ \frac \partial {\partial t}+v_1\frac \partial {\partial R}\right]
F_1=
\]
\begin{equation}
=\rho \int dp_2\int d^2B\left| v_1-v_2\right| \{F_2^{in}(t,R,p_1^{\prime
},p_2^{\prime })-F_2^{in}(t,R,p_1,p_2)\}\equiv S_{12}(F_2^{in})
\end{equation}
where the ansatz about molecular chaos (the famous Stosshalansatz) is
introduced,
\[
F_2^{in}(t,R,p_1,p_2)=F_1(t,r_1\approx R,p_1)F_1(t,r_2\approx R,p_2)
\]
Here $F_2^{in}(t,R,p_1,p_2)$ is DF of two particles which come into
collision at a point $R$ from the preceeding in-state, $B$ is the
two-dimensional impact parameter vector perpendicular to the relative
velocity $v=v_1-v_2$ ( $B=r-v(v,r)/\left| v\right| ^2$ with $r=r_1-r_2$ ),
and $p_1^{\prime },p_2^{\prime }$ are the momentums what should be
attributed to initial in-state in order to transform into the momentums
$p_1,p_2$ at final out-state after collision. In accordance with
Stosshalansatz, the join two-particle DF is factored into the product of
one-particle DFs.

\subsection{IS BOLTZMANN EQUATION DERIVABLE\\ FROM BBGKY HIERARCHY ?}

With no doubts, BE is the well grounded kinetical model. Thus it is the
striking surprise that it still is not derived from BBGKY equations.

The widely spread opinion is that the derivation was performed by
Bogolyubov. However, it was based on two ansatzes. First, the evolution of
statistical ensemble governed by BBGKY equations tends to the stage when all
the DFs become time-local functionals of one-particle DF. Secondly, the
molecular chaos (mistakenly identified wih the weakening of correlations of
far spaced particles).

Both these assumptions seem quite reasonable. But if it is really so, both
should be deducable from BBGKY formalism. The possibility of rigirous
derivation of BE from BBGKY equations was carefully analysed by O.Lanford,
under a suitable definition of the norm in DF's functional space in the
Boltzmann-Grad limit [35,36] (see also [30,37]). It was shown that the
divergency of the series expansion of solution of BBGKY hierarchy to
solution of BE takes place, but only during a short time after start smaller
than the free flight time.

The attempts to extend this result to larger time intervals run into likely
principal difficulties. The matter is that the singularities
(discontinuities) of binary DF and higher DFs arise at hypersurfaces in the
phase space which correspond to collisional configurations (for instance,
when the vector $v_2-v_1$ is parallel to $r_2-r_1$ ). Though these
singularities are concentrated in regions whose Lebesgue measure tends to
zero in Boltzmann-Grad limit, thats are just the regions which determine the
evolution of DFs of colliding particles and reliability of molecular chaos.

\subsection{OBJECTIONS TO MOLECULAR CHAOS\\ AND BOLTZMANN EQUATION}

M.Kac [31] had doubts that BE is derivable from the BBGKY theory in general
spatially non-uniform case. He treated BE sooner as the good model than as
formal approximation of exact theory. According to Kac, in the uniform
situation one could come to BE by means of spatial averaging in addition to
ensemble averaging. In this case DFs contain no information about previous
motion of gas particles, therefore the statistical ensemble makes no
difference between two actually colliding particles and two arbitrary
particles, and molecular chaos looks quite inevitable. But in general one
losses the help of spatial averaging.

This criticizm was developed in [9]. According to BE, a life of a gas
particle is prescribed by quite certain portions of collision of various
sorts (by the portioning what corresponds to uniform distribution of the
impact vector $B$ ). But in reality the distribution over sorts is out of
both dynamical and thermodynamical control of the system. Even if it
happened that some portion was much larger than its expected value, this
incident will not influence the future. There are no mechanism what would be
able to remember the history of preceding collisions and use it in order to
compensate previous occasions and keep a certain distribution (histogram) of
collisions over sorts.

Consequently, in accordance with the above phenomenology, the number of
collisions (of any given sort) per unit time may have no predictable limit.
In other words, the rate and the efficiency (the effective cross-section) of
the flow of collisions of a gas particle undergo scaleless flicker
fluctuations. The effect of these fluctuations on the evolution of $F_1$
could be diminished only by spatial averaging in the uniform gas, but in no
way in the framework of non-uniform ensemble, when the spatial dependencies
of DFs carry an information about history of Brownian motion of gas
particles [9].

Indeed, the rate (the diffusivity) of the Brownian motion is closely
correlated with fluctuations in the number and efficiency of collisions,
hence, with the currently possible collision too. Therefore, the present
collision has long-living statistical correlations with the past motion of
colliding particles. In view of these correlations, the binary DF for
particles going into collision acquires the specific statistical meanig as
the conditional probability distribution, under the condition that the joint
collision is realized in a fixed space-time point. Hence, this DF can not be
factored into product of DFs $F_1$.

The generally possible factoration has the form

\[
F_2^{in}(t,R,p_1,p_2)=F_1^{\prime }(t,R,p_1)F_1^{\prime \prime }(t,R,p_2)
\]
with conditional marginal distributions different from the conditionless
one-particle DF $F_1$, $F_1^{\prime }\neq F_1$, $F_1^{\prime \prime }\neq
F_1 $ , what corresponds to mutual independency of particles in the momentum
space but not in the real configurational space. Thus, the molecular chaos
fails, at least because of spatial statistical correlation of colliding
particles.

Such the ensemble can not be reduced to BE. It is amusing that the violation
of molecular chaos takes place only between colliding particles, that is
just where the molecular chaos is especially believed. This picture is
obviously incompatible also with the ansatzes used by Bogolyubov.

Therefore, we must introduce a different ansatz, with the same purpose to
build a kinetical model and firstly to narrow a variety of kinetically
describable solutions of BBGKY equations.

\section{CORRECT REFORMULATION\\ OF BOLTZMANNIAN KINETICS}

\subsection{PRIMARY ANSATZ OF GAS KINETICS}

The essence of coarsened kinetical language is that all the set of
successive dynamical stages of a two-particle interaction process is treated
as the unique momentary collision event. This event looks as a ''black box''
(compressed into point under Boltzmann-Grad limit) which instantly
transforms the input velocities into output ones.

Clearly, this collision event should conserve probabilities, else a model
would effectively allow for creation or annihilation of particles inside the
black box. In other words, in any kinetical statistical ensemble, the flow
of particles entering the collision should currently coinside with the flow
of particles what leave it.

In our opinion, just this trivial requirement should serve as the basic
ansatz when deriving a kinetical model from BBGKY equations. In order to
provide with the probability conservation, the exclusion of internal
dynamics of collision from consideration must be compensated by definite
bondary conditions at the black box borders. Concretely, one should keep the
probability measure of any final output state at the border to be currently
equal to the probability of corresponding initial input state.

This condition is formally similar to the one for DFs at $\left|
r_j-r_k\right| =d$ in the rigid sphere gas (with $d$ being the diameter of
spheres). In general, a choise of the borders of collision box is rather
arbitrary, at least in the framework of Boltzmann-Grad limit. Therefore the
above mentioned ansatz is equivalent to that probabilities of all the inner
stages of any fixed sort of collision equal to one and the same quantity. If
this condition was not satisfied, it would be impossible to use the concept
of momentary collision event at all!

In such the way we narrow the set of solutions of BBGKY hierarchy to be
under kinetical consideration: the distributions somehow modulated along a
path inside collision box are excluded (although arbitrary modulation from
one path to another is permissible).

The full characterization of a sort of binary collision consists of its
location defined by three space coordinates, the two-dimensional impact
parameter vector and two velocities of particles, that is totally by
3+2+6=11 scalars instead of total 12 phase space variables of the pair. The
rest twelfth variable, let be denoted as $\Theta $ , just enumerates inner
dynamical stages of the collision in the black box interior. This variable
should be excluded from a kinetical picture. Thus, with respect to $F_2$ ,
the primary ansatz means that
\begin{equation}
\frac \partial {\partial \Theta }F_2=0\,\,,\,\,r_2-r_1\in collision\,box
\end{equation}

One always can choose $\Theta $ as the inner time of the collision, for
instance, as the time necessary to achieve its given stage from its most
proximity stage. Then the differentiation by $\Theta $ looks as
\begin{equation}
\frac \partial {\partial \Theta }=v\frac \partial {\partial r}+f(r)(\frac
\partial {\partial p_2}-\frac \partial {\partial p_1})=v\frac \partial
{\partial r}+\Lambda _{21}\equiv L_{rel}^{(2)}
\end{equation}
with $r=r_2-r_1$ and $v=v_2-v_1$ being the relative distance and velocity of
particles, respectively. The operator $L_{rel}^{(2)}$ is nothing but the
part of full two-particle Liouville operator $L^{(2)}$ responsible for the
inner dynamics of collision.

Of course, none of rigorous solutions on BBGKY equations exactly satisfies
(2), as well as the ansatzes introduced by Boltzmann and Bogolyubov. But if
the kinetic approximation is actually available, we may hope that at a more
or less late stage (kinetical stage) of the ensemble evolution the ansatz
(2) is quantitatively satisfied to the extent sufficient to regard it as
formal identity.

\subsection{COLLISION INTEGRAL}

In fact, one inevitably needs in the primary ansatz in order to construct
the Boltzmannian collision integral, even under the molecular chaos
postulate, and thus to reduce the explicite dynamics to the momentary
collisions representation. Indeed, let us rewrite the right-hand side of
first BBGKY equation as

\[
I_{12}(F_2)=\rho \int \Lambda _{12}F_2dp_2dr_2=
\]
\[
=\rho \int \int_{CB}(v_2-v_1)\frac \partial {\partial r}F_2drdp_2-\rho \int
\int_{CB}\frac \partial {\partial \Theta }F_2drdp_2
\]
where $CB$ marks the spatial integration over the collision black box. After
transformation the volume integral in the first right-hand term into surface
integral $\oint $ over the collision box board and after separation of
in-states and out-states, regarding of sign of the scalar product $(vr)$ ,
one obtains
\[
\ I_{12}(F_2)=\rho \int \left[ \oint_{vr>0}\left| v\right|
F_2^{out}d^2B^{\prime }-\oint_{vr<0}\left| v\right| F_2^{in}d^2B-
\int_{CB} \frac{\partial F_2}{\partial \Theta }dr\right] dp_2
\]
Here in both the surface integrals one and the same velocity arguments are
present while the impact parameter vectors $B^{\prime }$ and $B$ relate to
final and initial state, respectively.

This expression can be reduced to conventional collision integral only if
the probabilities of the final states constantly coinside with thats of
preceding incoming states,
\[
F_2^{out}(t,R,p_1,p_2,B^{\prime })=F_2^{in}(t,R,p_1^{\prime },p_2^{\prime
},B)\
\]
Here $R$ is the location of collision, and in view of Boltzmann-Grad limit
we do not make difference between spatial locations of in-states and
out-states.

The validity of this relation is ensured just by ansatz (2). Besides, due to
(2) the last integral, what formally acts like a bulk source of particles,
turns into zero. As the result, we come to the kinetic equation (1), but, as
it will be seen, the molecular chaos hypothesis fails.

\subsection{VIOLATION OF MOLECULAR CHAOS}

According to primary ansatz, in the interior of collision box the second
BBGKY equation reduces to
\begin{equation}
\left[ \frac \partial {\partial t}+U_2\frac \partial {\partial R}+\frac
\partial {\partial \Theta }\right] F_2\Rightarrow \left[ \frac \partial
{\partial t}+U_2\frac \partial {\partial R}\right]
F_2=I_{13}(F_3)+I_{23}(F_3)
\end{equation}
with $R\equiv (r_1+r_2)/2$ and $U_2\equiv (v_1+v_2)/2$ being the position
and velocity of the mass center of colliding pair, i.e. the position and
velocity of the collision as a whole.

Hence, the spatial drift of the DF of colliding or closely spaced particles
is governed by the mass center velocity. This consequence of (2) has very
simple physical meaning. First, if the relative motion of particles is
included into the collision interior, then quite naturally it should be
automatically excluded from the transport of the collision as a whole.
Secondly, the DF related to collision box represents the conditional
probability under the condition that the particles are partners in one and
the same encounter. It is natural that this specified DF is ruled by an
equation different from what describes the less informative unconditional
DF. Evidently, the conditional DF $F_2$ in (4) can be interpreted as the
probability measure of that a collision occures in the point $R$ , i.e. as
the measure of ensemble-averaged density of collisions.

It is easy to see that in a spatially non-uniform ensemble, when $\partial
F_2/\partial R\neq 0$ , the Eq.4 forbids the factoration of this DF into
product of one-particles DFs, $F_2\neq F_1(t,p_1,R)*F_1(t,p_2,R)$ ,
regardless of contribution of the right-hand side. Therefore the molecular
chaos inevitably fails, and just with respect to colliding particles! As a
consequence, in general the density of collisions is not simply proportional
to [density of particles]$^2$ .

\subsection{3-PARTICLE PROCESSES AND EVOLUTION\\ OF DENSITY OF COLLISIONS
}

Let us take in mind the Boltzmann-Grad limit (BGL),
\[
\rho \rightarrow \infty ,\,d\rightarrow 0,\,\lambda =1/\rho d^2=const
\]
where $\lambda $ is the mean free path and $d$ is characteristic radius of
interaction.

The right-hand side of the Eq.4 describes a change of density of collisions
due to three-particle processes. Thats are not factual triple collisions,
which are of no importance under BGL, but two successive pair collisions
separated by a distance comparable with $d$ but perhaply much greater
than $d $ . One of the collisions involves a third external
particle representing
the rest gas. Under BGL, a spatial region taken by this process contracts
into point if measured by $\lambda $ scale although can grow up to infinity
in $d$ units.

In order to express the right-hand side of Eq.4 in kinetic fashion, in terms
of collision integrals, we need in the obvious three-partricle extention of
the ansatz (2),

\[
\partial F_3/\partial \Theta =L_{rel}^{(3)}F_3\Rightarrow
0\,\,,\,L_{rel}^{(3)}=L^{(3)}-U_3\partial /\partial R\,
\]
where $\Theta $ is a suitably choosen inner time of the process, and $R$ and
$U_3$ are the position and velocity of mass center of the close
three-particle configuration under consideration. Eventually, the Eq.4 is
transformed into the kinetical equation

\begin{equation}
\left[ \frac \partial {\partial t}+U_2\frac \partial {\partial R}\right]
F_2=S_{13}(F_3^{in})+S_{23}(F_3^{in})
\end{equation}
where the superscript $in$ means that the external third particle is at
in-state with respect to 1-st and 2-nd particles (i.e. is coming to them
from the infinity, in $d$ units). This equation describes the evolution of
density of binary encounters and collisions as changed by collisions of the
given two particles with the rest gas.

\subsection{NO FAR CORRELATIONS\\ ALTHOUGH NO MOLECULAR CHAOS}

In contrary to the conditional binary DF subjected to Eq.4, the general
binary DF for arbitrary configurations placed outside the collision box is
subjected to the usual equation
\begin{equation}
\left[ \frac \partial {\partial t}+v_1\frac \partial {\partial r_1}+v_2\frac
\partial {\partial r_2}\right] F_2=S_{13}(F_3^{in})+S_{23}(F_3^{in})
\end{equation}
Evidently, this equation allows for the factored solutions,
$F_2=F_1(t,r_1,p_1)*F_1(t,r_2,p_2)$ ,
with $F_3^{in}=F_2^{in}(t,r_1,p_1,p_3,B)*F_1(t,r_2,p_2)$ in
$S_{13}(F_3^{in})$ and
$F_3^{in}=F_2^{in}(t,r_2,p_2,p_3,B)*F_1(t,r_1,p_1)$ in $S_{23}(F_3^{in})$ ,
where each of $F_1$ multipliers is the solution on the Eq.1.

Thus the vanishing of correlations of far distanced particles coexists with
invalidity of molecular chaos with respect to closely spaced particles (in
particular, colliding ones). If measure the inter-particle distance with the
free flight path $\lambda $ then, under BGL, the weakening of
correlations, $F_2=F_1(t,r_1,p_1)*F_1(t,r_2,p_2)$ , can be rigorously
satisfied anywhere
except the hypersurface $r_2-r_1=0$ . In view of this circumstance, the
singularities of DFs at collisional hypersurfaces met in [35,36] confirm the
necessity to specially consider the specific DFs for encountering
configurations with $r_1=..=r_n$ .

\subsection{MANY-PARTICLE ENCOUNTERS AND\\ CHAIN OF KINETICAL EQUATIONS}

In the framework of BGL, one should continue the above mentioned procedure
up to infinity, with extention of primary ansatz to many-particle events,
namely, to sequences of pair collisions and close many-particle encounters,
step by step building the kinetical model of spatially non-uniform gas. The
extention looks as $\partial F_n/\partial \Theta =0$ , with the equality to
be satisfied inside a collision (encounter) black box alloted for the event
and $\Theta $ being its inner time. This is the inevitable payment for
decision to reformulate the ensemble evolution in terms of collision
integrals.

Because of identities $\partial /\partial \Theta =L_{rel}^{(n)}$ ,
$L^{(n)}=U_n\partial /\partial R+L_{rel}^{(n)}$ , where $L_{rel}^{(n)}$ is
the part of $n$-particle Liouville operator responsible for relative motion
inside the event, the result is the infinite chain of coupled kinetic
equations [9]
\begin{equation}
\left[ \frac \partial {\partial t}+U_n\frac \partial {\partial R}\right]
F_n=\sum_{j=1}^nS_{jn+1}(F_{n+1}^{in})
\end{equation}
Here $R=(r_1+..+r_n)/n$ , $U_n=(v_1+..+v_n)/n$ .

The DFs on the left represent the probability measure of occurence
of $n$-particle encounters, in particular, constituted
by $(n-1)$ successive
collisions. The right-hand DFs relate to $(n+1)$-particles events, with an
external $(n+1)$-th particle being at precollisional in-state with respect
to the left $n$-particle configuration. In general, any of these $F_n$
depends on time, the location of the encounter $R$ , $n$ momentums, $(n-1)$
impact parameter vectors and, besides, on $(n-2)$ distancies between binary
encounters if $n>2$ . Again, the relative motion turns into the hidden inner
property of the event and so does not influence the drift of the probability
of the event as a whole.

In spatially uniform case, when $\partial /\partial R\rightarrow 0$ , the
solution on the Eqs.7 can be sought in the conventional molecular chaos
form, $F_n=\prod_jF_1(t,r_j,p_j)$ , with no factual dependence on impact
parameters, and the model reduces to conventional BE.

\subsection{MORE ABOUT PRIMARY ANSATZ}

To formulate the primary ansatz specially for BGL, let us write the
interaction force in the form $f(r)\Rightarrow f(r/d)/d$ and measure the
position of a cluster of close particles in $\lambda $ units (and time in
free flight time units) while interparticle distancies in $d$ units. Then
BBGKY equations turn into
\[
\left[ \frac \partial {\partial t}+U_n\frac \partial {\partial R}+\frac 1\mu
L_{rel}^{(n)}\right] F_n^{(\mu )}=\int \sum_j\Lambda _{jn+1}F_{n+1}^{(\mu
)}dp_{n+1}dr_{n+1}
\]
where $\mu =d/\lambda =\rho d^3\rightarrow 0$ is small gas parameter, and
DFs dependence on $\mu $ is marked. Clearly, if a definite limit of all
the $F_n^{(\mu )}$ does exist at $\mu \rightarrow 0$ , then without fail the
relation $\partial F_n^{(0)}/\partial \Theta =L_{rel}^{(n)}F_n^{(0)}=0$ is
satisfied. Hence, the ansatz is equivalent to the assumption that
$L_{rel}^{(n)}F_n^{(\mu )}=o(\mu )$ , at least for configurations
corresponding to chains of far spaced binary collisions (i.e excluding
actually triple or multiple collisions).

\subsection{COLLISIONS, ENCOUNTERS AND\\ COARSENED KINETIC EQUATIONS}

In principle, under a suitable geometrical parametrization of the events the
left-hand DFs in Eqs.7 can be identified with the right-hand ones. However,
a full classification of configurations would lead to an extremely
complicated theory.

It seems reasonable to simplify the model with the help of spatial averaging
of DFs over $r_j$ inside the collision (encounter) regions, under fixed mass
center position $R$ . Of course, then a dependence on impact parameters will
be lost, except the only parameter on the right-hand sides what relates to
the incoming $(n+1)$-th external particle. As a consequence, one will be
forced to distinguish left-hand DFs $F_n$ and right-hand DFs $F_n^{in}$ .

Besides, for more simplicity, in the framework of BGL we may choose the
dimensions of the collision (encounter) regions (boxes) much larger than
interaction diameter $d$ . Then the most part of configurations under
averaging will correspond to encounters, when particles are separated by a
distance comparable with $d$ but without factual interaction (and so
out-states coinside with in-states).

We may hope that in spite of such the crucial simplification the essential
features of statistics of collisions will not be lost. Indeed, in view of
incident origin of close $n$-particle clusters, there are no principal
difference between efficient collisions and merely encounters. The latters
can be regarded as collisions with negligibly weak scattering, and
fluctuations in the flow of collisions should be similar to thats in the
flow of encounters.

\subsection{WEAKENED MOLECULAR CHAOS}

It is interesting that the chain (7) is still formally time-reversible if
all the geometry of binary and many-particle events is kept under control.
Indeed, the combined inversion of time and velocities and rearrangement of
in-states and out-states does not change these equations. However, this does
not prevent existence of seemingly irreversible solutions of the equations,
all the more if a part of the complete characterization of many-particle
configurations is lost.

The DFs smoothed over collision regions have the clear physical meaning:
thats represent ensemble-averaged densities of $n$-particle encounters, each
depending on time, location $R$ and $n$ velocities. But in order to close
the Eqs.7 one needs in some connections between $F_{n+1}^{in}$ and the
left-side functions. This is just the point where one more ansatz like
molecular chaos should be attracted.

It is natural to assume that the velocity distribution of $(n+1)$-th
right-hand external particle is statistically independent on velocities of
the left-hand cluster particles:

\[
F_{n+1}^{in}(t,R,p^{(n+1)})=G_n(t,R,p^{(n)})F_1(t,R,p_{n+1})
\]
where the notation $p^{(n)}\equiv \{p_1,..,p_n\}$ is introduced. In general
this factoration is not identical to absolute statistical independence of
particles, since it allows for spatial correlations whose manifestation is
that functions $G_n$ differ from $F_n$ .

Clearly, one should connect $F_{n+1}^{in}$and thus $G_n$ with $F_{n+1}$ ,
not with $F_n$ . The matter is that the spatial dependencies of both
$F_{n+1}^{in}$and $F_{n+1}$ reflect the probability density of
the same $(n+1) $-particle encounter events. Therefore the degree of
specifically
spatial correlations must be one and the same in both
$F_{n+1}^{in}$and $F_{n+1}$ (this assertion follows also merely from
the proximity of $(n+1)$-th particle to the left-side cluster in
$\lambda $ units). This requirement
implies

\[
\int F_{n+1}dp^{(n+1)}=\int F_{n+1}^{in}dp^{(n+1)}\Rightarrow \int
G_n(t,R,p^{(n)})dp^{(n)}\int F_1(t,R,p^{\prime })dp^{\prime }
\]
One can say also that this relation expresses the conservation of particles
and probabilities during encounters.

Besides, take into account that the degree of correlations inside the
left-hand cluster must be the same in $F_{n+1}^{in}$and in $F_{n+1}$. Then
the only reasonable form of $G_n$ is $G_n=\int F_{n+1}dp_{n+1}/\int F_1dp_1$,
and we finally come to the connection [9]
\begin{equation}
F_{n+1}^{in}(t,R,p^{(n+1)})=\int F_{n+1}(t,R,p^{(n)},p^{\prime })dp^{\prime }
\frac{F_1(t,R,p_{n+1})}{\int F_1(t,R,p^{\prime })dp^{\prime }}
\end{equation}

This is the weakened form of molecular chaos which establishes the
statistical independence of the external particle only in the velocity
space, but not in the configurational space.

\subsection{NON-UNIFORM EQUILIBRIUM GAS\\ AND BOLTZMANN-LORENTZ MODEL}

One naturally is forced to consider some spatially non-uniform solutions on
BBGKY hierachy even in thermodynamically equilibrium case, if wants to
analyse the statistics of Brownian motion of a test gas particle. If the
position of even a single test particle is known at initial time moment, or
even merely its distribution differs from the uniform one, then one should
deal with spatially varying DFs.

Let DF $F_1(t,R,p_1)$ be related to the test particle. Then the higher order
DFs $F_{n+1}$ subjected to Eqs.7 and 8 relate to the clusters consisting of
the test particle plus $n$ arbitrary close particles. All the DFs together
encode the information about statistics of encounters, collisions and random
displacements of test particle.

But, of course, now the DF of test particle, $F_1(t,R,p_1)$ , and the
one-particle DF $F_1(t,R,p_{n+1})$ in (8) are different things. The latter
relates to arbitrary gas particle and so should coinside with the Maxwellian
distribution in equilibrium gas:
\[
F_1(t,R,p_{n+1})=F^{eq}(p_{n+1})\,\,,\,\,F^{eq}(p)\equiv (2\pi
Tm)^{-3/2}\exp (-p^2/2mT)
\]
The weakened molecular chaos condition transforms into
\begin{equation}
F_{n+1}^{in}(t,R,p^{(n)},p_{n+1})=F^{eq}(p_{n+1})\int
F_{n+1}(t,R,p^{(n)},p^{\prime })dp^{\prime }
\end{equation}
The sets of equations (7) and (9) represent the analogy of the well known
Boltzmann-Lorentz model but correctly modified for a spatially non-uniform
ensemble.

\section{SELF-DIFFUSION AND\\ DIFFUSIVITY 1/F-NOISE}

\subsection{KINETICAL EQUATIONS FOR BROWNIAN\\ MOTION OF TEST PARTICLE}

Let $r(t)$ and $u(t)$ be coordinate and velocity of any particle and
$Q(t)=\int_0^tu(t^{\prime })dt^{\prime }$ be its displacement during time
interval $t$ (more accurately, let $u$ and $Q(t)$ be the projections of
velocity and displacement vector onto a fixed direction). The statistics of
self-diffusion (Brownian motion) of gas particle is described by the
characteristic function (CF), $\left\langle \exp [ikQ(t)]\right\rangle $ ,
but, as we know, the most significant information is contained in the first
two terms of expansion of $\ln \left\langle \exp [ikQ(t)]\right\rangle $
over $k^2$ , which reflect the mean diffusivity and the correlation function
of diffusivity fluctuations.

The Laplace transform of CF always can be represented as
\[
\int_0^\infty \left\langle \exp [ikQ(t)]\right\rangle \exp
(-zt)dt=[z+kD(z,ik)k]^{-1}
\]
with $D(z,ik)$ being the frequency dependent diffusivity kernel. From the
other hand, with the help of generalized Green-Kubo formulas, one can
connect $D(z,ik)$ and the response of probability distribution of a test
particle to a weak external force $f_{ext}(r)$ , $f_{ext}\rightarrow 0$ ,
which influence it after switching on at $t=0$ (being parallel to the
choosen direction). Concretely [9],
\[
\int_0^\infty dt\exp (-zt)\int dr\exp (-ikr)\int vF_1(t,r,p)dp=
\]

\[
=\frac{D(z,ik)}{T[z+kD(z,ik)k]}\int \exp (-ikr)f_{ext}(r)dr
\]
where $v=p/m$ , the DF $F_1(t,r,p)$ describes the test particle and equals
to $F^{eq}(p)$ at $t<0$ .

Hence, the long-time statistics of equilibrium self-diffusion can be
obtained from the spatially non-uniform evolution (formally nonequilibrium
but in fact equilibrium in the macroscopic sense) of ensemble disturbed by a
force what acts on the only test particle. In the framework of the kinetical
theory constructed, the problem reduces to solution of Eqs.7 and 9 , after
adding the terms $f_{ext}(R)\partial F_n/\partial p_1$ to all left sides of
Eqs.7.

We may avoid a part of non-principal mathematical difficulties if replace
the Boltzmannian collision integrals on right-hand sides of Eqs.7 with the
Fokker-Planck approximation of the Boltzmann-Lorentz collision operator,

\[
S_{jn+1}(\int F_{n+1}(t,R,p^{(n)},p^{\prime })dp^{\prime
}*F^{eq}(p_{n+1}))\Rightarrow \gamma \frac \partial {\partial
p_j}(p_j+mT\frac \partial {\partial p_j})\int F_{n+1}dp_{n+1}
\]
with $\gamma $ being the inverse free flight time. Then we have to solve the
equations [9]
\begin{equation}
\left[ \frac \partial {\partial t}+U_n\frac \partial {\partial
R}+f_{ext}(R)\frac \partial {\partial p_1}\right] F_n=\sum_{j=1}^n\gamma
\frac \partial {\partial p_j}(p_j+mT\frac \partial {\partial p_j})\int
F_{n+1}dp_{n+1}
\end{equation}
with $U_n=(v_1+..+v_n)/n$ and equilibrium initial conditions $F_n\mid
_{t=0}=\prod_jF^{eq}(p_j)$ .

\subsection{NON-LOCAL DIFFUSIVITY KERNEL}

But even the analysis of these simplific equations is rather hard, and only
the lowest terms of the expansion
\[
D(z,ik)=D_0(z)+(ik)^2D_1(z)+...
\]
were found in [9]. Nevertheless, these terms always allow to check the
existence of low-frequency duffusivity fluctuations and estimate their
variance.

In accordance with the $D(z,ik)$ definition, one can verify that $D_0(z)$ is
the Laplace transform of the two-point equilibrium velocity correlator,
$K_u(t)\equiv \left\langle u(t)u(0)\right\rangle $ , and
\[
D_1(z)=\int_0^\infty \{K_D(t)+\,K_u(t)\int_0^t(t-t^{\prime })K_u(t^{\prime
})dt^{\prime }\}\exp (-zt)dt
\]
Here $K_D(t)$ is the above discussed diffusivity correlation function as
expressed via four-point equilibrium cumulant of $u(t)$ . Therefore, at low
frequencies, $\left| z\right| <<\gamma $ , $D_1(z)$ is nothing but the
Laplace transform of $K_D(t)$ , that is $Re[D_1(2\pi if+0)]$ is responsible
for the power spectrum of diffusivity fluctuations. From the other hand, one
can connect $D_1(z)$ strictly with the fourth statistical moment of the
displacement:
\[
\int_0^\infty \left\langle Q^4(t)\right\rangle \exp (-zt)dt\,\,=
\frac{24}{z^2 }[D_1(z)+D_0^2(z)/z]
\]

The latter function was obtained in [9] in the form of double series
via $\gamma /z$ which can be exactly transformed into an integral.
Here we omit
these manipulations. At $\left| z\right| <<\gamma $ , the final result reeds
as
\[
D_1(z)\approx \frac 1{2z}D_0^2(z)[\ln ^2\frac \gamma z+const]
\]
with $D_0(z)=\frac Tm(\gamma +z)^{-1}\,\,$and $D_0=D_0(0)=T/m\gamma $ being
the usual diffusivity. In the temporal reprezentation, these results mean
that $\left\langle Q^2(t)\right\rangle =2D_0t$ and
\[
\left\langle Q^{(4)}(t)\right\rangle =\left\langle Q^4(t)\right\rangle
-3\left\langle Q^2(t)\right\rangle ^2\approx 6D_0^2t^2[\ln ^2\gamma
t+c^{\prime }\ln \gamma t+c^{\prime \prime }]
\]
at $\gamma t>>1$ , where $c^{\prime },c^{\prime \prime }$ are numerical
constants. In view of the above consideration, this is the clear evidence of
existence of 1/f type self-diffusivity fluctuations.

Formally, these results are produced by the peculiar drift terms $U_n\frac
\partial {\partial R}$ on right-hand sides of the Eqs.10. But our experience
showed the remarkable fact that the concrete form $U_n=(v_1+..+v_n)/n$ is of
no principal importance. Almost any sequence of $U_n=\sum c_{nj}v_j$ with
different $c_{nj}$ results in an excess contribution to the fourth cumulant
$\left\langle Q^{(4)}(t)\right\rangle $ which grows faster than
proportionally to time and thus corresponds to a low-frequency diffusivity
noise. In particular, if the mass of the test particle $M$ differs from
masses of gas particles, and so $U_n=(Mv_1+mv_2+...)/(M+nm)$ , then the
Eqs.10 lead to $\left\langle Q^{(4)}(t)\right\rangle \sim t^{1+\gamma }$ ,
$\gamma >0$ , where $\gamma $ differs from unit and depends on
the ratio $M/m$ [9].

\subsection{1/F-FLUCTUATIONS OF DIFFUSIVITY}

The power spectrum of the relative diffusivity fluctuations resulting from
the obtained $D_1(z)$ for sufficiently low frequences, $f<<\gamma $ , is
\[
S_{\delta D}(f)\approx \frac 1{2f}\ln \frac \gamma f
\]
Formally this is non-integrable spectrum like that of a slightly
non-stationary noise. But we already emphasized that it is fictituous
non-stationarity what refers only to the duration of observation, not to the
time moment when observation takes beginning. Besides, it is necessary to
underline that the spectrum is insensitive to spatial dimensionality of gas,
giving the illustration of so-called zero dimensionality of 1/f-noise.

\section{RESUME}

We reviewed the explanation of 1/f-noise based on the idea that its sources
are not some specifically slow processes but, in opposite, the same
processes which realize forgetting the past, relaxation and related white
noise. If a system producing randomness constantly forgets its past and
keeps in its memory scope only a limited amount of recently happened events,
then it is unable to definitely distinguish a proper part of events and
deviation from it. As a consequence, the fluctuations in the amount of
incidently emerging events grow similarly to its average value, resulting in
just 1/f-fluctuations of the rate of random flow of events.

In order to comprehend such a random behaviour, one should make difference
between long-living statistical correlations and long-living causal
correlations.

We noted that 1/f-noises in different systems can be interpreted in such
manner. The corresponding phenomenological models of purely random flow of
events and of Brownian motion were considered, whose important feature is
violation of the central limit theorem with respect to long-range statistics
and close connections with Cauchy statistics.

The attempts to develope an adequate philosophy of 1/f-noise were made. The
quantitative estimates of 1/f-noise in terms of only short-range
(microscopic) characteristic scales were presented, showing that the wider
is the memory scope of the system the lower is 1/f-noise level.

For thermodynamical systems, the general connections between 1/f-noise and
four-point equilibrium cumulants were analysed. The existence of long-living
four-point correlations equivalent to flicker noise in exactly solvable
Kac's ring model was found.

The connections with Hamiltonian models of relaxation and irreversible
processes were discussed, with hopes that if any realistic time-reversible
many-particles dynamics produces noise and irreversibility then it also
produces 1/f-fluctuations of their rates. This expectation can be testified
in such principally important cases as slightly unharmonic crystal lattice
(phononic system) and classical gas with short repulsive interactions (in
the Boltzmann-Grad limit).

We pointed out that collisions of any gas particle with other particles form
just such a random flow of events whose rate is being constantly forgotten
and free of a thermodynamical control by the system. We argued that
consequently 1/f fluctuations of the rate (number of collisions per unit
time) and of related quantities (effective cross-section, self-diffusuvity,
mobility) should occur and manifest themselves as the spatial statistical
correlations of colliding particles in a spatially non-uniform statistical
ensemble, resulting in violation of both the molecular chaos ansatz and the
ansatz used by Bogolyubov to deduce Boltzmann equation from BBGKY hierarchy.

The euristic reasonings were confirmed by critical analysis of kinetical
theory of gas. It was argued that reformulation of the gas ensemble
evolution in terms of coarsened description with Boltzmannian collision
integrals inevitably needs in another ansatz whose meaning is the
conservation of particles and constancy of probabilities on the path what
connects in-state and out-state through the collision box (i.e. the
space-time region to be excluded from kinetical description).

We showed that this requirement leads to the infinite set of specific
kinetical equations for the probability distributions on the hypersurfaces
which correspond to binary and many-particle encounters and collisions (the
sequences of binary collisions, not actually multiple ones). In a spatially
non-uiform ensemble these kinetical equations forbid conventional molecular
chaos, because of producing statistical correlations of close particles in
the configuration space, and can be reduced to standard Boltzmann equation
only in the limit uniform case.

In our opinion, this is in agreement with the results by O.Lanford who noted
the appearance of singularities of distribution functions at the above
mentioned hypersurfaces in the formal series solution of BBGKY equations
when investigating the possibility of reduction of BBGKY hierarchy to
Boltzmann equation. The singularities just indicate that one needs in
special consideration of probabilities of colliding and close configurations.

The obtained chain of kinetical equations, with addition Green-Kubo formulas
and higher-order fluctuation-dissipation relations, was applied to analyse
the statistics of self-diffusion (Brownian motion of a test particle) in
equilibrium gas. The four-order cumulant of Brownian displacement was
calculated whose time behaviour proves the existence of 1/f-fluctuations of
diffusivities (and thus of mobilities) of gas particles.

Hence, we have at least one example of principally correct formulation of
kinetical approximation which does not loss 1/f-noise, and the example of
dynamical system by no means producing slow processes but in spite of this
producing 1/f-noise.

\thinspace {\bf ACKNOWLEDGEMENTS}

I would like to acknowledge the participants of the seminars in the
Department of kinetic properties of disordered systems in Donetsk
Phys.-Techn. Institute of NASU, in the Nijnniy-Novgorod (Gorkii) State
University in Russia and in the Moscow State University, and in the
International Solvey Institutes for Physics and Chemistry in Brussels for
interesting discussions.

\thinspace \thinspace

{\bf REFERENCES}

1. P.Dutta, P.Horn, ''Low-frequency fluctuations in solids: 1/f noise'',
Rev.Mod.Phys., 53(3), 497 (1981).

2. F.N.Hooge, T.G.M.Kleinpenning, L.K.J.Vandamme, ''Experimental studies of
1/f noise'', Rep.Prog.Phys., 44, 481 (1981).

3. N.S.Krylov, ''Works on the foundations of statistical mechanics'',
Princeton U.P., Princeton, 1979 (the original books published in Russian in
1950).

4. Yu.E.Kuzovlev and G.N.Bochkov. ''On nature and statistics of 1/f-noise''.
Preprint No.157, NIRFI (Scientific-Research Radiophysics Institute), Gorkii,
USSR, 1982 (in Russian).

5. Yu.E.Kuzovlev and G.N.Bochkov. ''On the origin and statistical
characteristics of 1/f-noise''. Izvestiya VUZov.-Radiofizika, 26, 310
(1983). Transl. in English in ''Radiophysics and Quantum Electronics''
(RPQEAC, USA), No 3 (1983).

6. G.N.Bochkov and Yu.E.Kuzovlev. ''On the probabilistic characteristics of
1/f-noise''. Izvestiya VUZov.-Radiofizika, 27, 1151 (1984). Transl. in
English in ''Radiophysics and Quantum Electronics'' (RPQEAC, USA), No 9
(1984).

7. G.N.Bochkov and Yu.E.Kuzovlev. ''New in 1/f-noise studies''. Uspekhi
fizicheskikh nauk, 141, 151 (1983). English translation: Sov.Phys.-Usp., 26,
829 (1983).

8. G.N.Bochkov and Yu.E.Kuzovlev. ''On the theory of 1/f-noise''. Preprint
No.195, NIRFI\ (Scientific-Research Radiophysics Institute), Gorkii, USSR,
1985 (in Russian).

9. Yu.E.Kuzovlev. ''Bogolyubov-Born-Green-Kirkwood-Yvon equations,
self-diffusion and 1/f-noise in a slightly nonideal gas''. Zh. Eksp. Teor.
Fiz, 94, No.12, 140 (1988). English translation: Sov.Phys.JETP, 67(12), 2469
(1988).

10. Yu.E.Kuzovlev, ''Purely random time partitions and 1/f-noise''.
Phys.Lett., A 194, 285 (1994).

11. Yu.E.Kuzovlev. ''Relaxation and 1/f-noise in phonon systems''. Zh. Eksp.
Teor. Fiz, 111, No.6, 2086 (1997). English translation: JETP, 84(6), 1138
(1997)

12. V.Feller, ''Introduction to probability theory and its applications'',
Mir, Moscow, 1967.

13. D.S.Ornstein, ''Ergodic theory, randomness and dynamical systems'', New
Haven, London, Yale University Press, 1974.

14. P.H.Handel, ''Nature of 1/f frequency fluctuations in quartz crystal
resonators'', Sol.-St. El., 22, 875 (1979).

15.Y.Nogushi, Y.Teramashi and T.Musha, ''1/f frequency fluctuations of a
quartz oscillator'', Appl.Phys.Lett., 40, 872 (1982).

16. T.Musha, B.Gabor and M.Shoji, ''1/f photon-number fluctuations in quartz
observed by light scattering'', Phys.Rev.Lett., 64(20), 2394 (1990).

17. R.F.Voss, ''Evolution of long-range fractal correlations and 1/f noise
in DNA base sequences'', Phys.Rev.Lett., 68, 3805 (1992).

18. W.Li, T.G.Marr and K.Kaneko, ''Understanding long-range correlations in
DNA sequences'', Physica, D 75, 392 (1994).

19. R.F.Voss and J.Clarke, Phys.Rev., B 13, 556, 1976.

20. M.Nelkin and A.M.S.Tremblay, ''Deviation of 1/f voltage fluctuations
from scale-similar Gaussian behaviour'', J.Stat.Phys., 25, 253 (1981).

21. J.L.Tandon and H.P.Bilger, ''1/f-noise as a non-stationary process:
evidences and some analytical conditions'', J.Appl.Phys., 47, 1697 (1976).

22. J.J.Brophy, ''Statistics of 1/f-noise'', Phys.Rev., 166(3), 827 (1968).

23. J.J.Brophy, ''Low-frequency variance noise'', J.Appl.Phys., 41, 2913
(1970).

24. A.A.Aleksandrov, G.N.Bochkov, A.A.Dubkov and A.I.Chikin, ''Measuring
probability distribution of 1/f-noise'', Izvestiya VUZov.-Radiofizika, 29,
980 (1986). Transl. in English in ''Radiophysics and Quantum Electronics''
(USA).

25. A.Sadbery, ''Quantum mechanics and the particles of nature'', Cambridge
Univ. Press, 1986.

26. I.Prigogine, ''From being to becoming'', W.H.Freeman and Co., San
Francisco, 1980.

27. R.D.Hughes, E.W.Montroll and M.F.Shlesinger, J.Stat.Phys., 28, 111
(1982).

28. I.Procaccia and H.Schuster, ''Functional renormalization-group theory of
universal 1/f noise in dynamical systems'', Phys.Rev., A 28, 1210 (1983).

29. D.V.Anosov et al., ''Dynamical systems with hyperbolic behaviour'', in
''Modern problems of mathematics. Fundamentals. Vol. 66'', ed.
R.V.Gamkrelidze, VINITI, Moscow, 1991 (in Russian).

30. G.Gallavotti, ''Statistical mechanics'', 1998.

31. M.Kac, ''Probability and related topics in physical sciences'',
Intersci. Publ., London, N.-Y., 1957.

32. G.N.Bochkov and Yu.E.Kuzovlev, ''On general theory of thermal
fluctuations in nonlinear systems'', Sov.Phys.-JETP, 45, 125 (1977).

33. G.N.Bochkov and Yu.E.Kuzovlev, ''Fluctuation-dissipation relations for
nonequilibrium processes in open systems'', Sov.Phys.-JETP, 49, 543 (1979).

34. G.N.Bochkov and Yu.E.Kuzovlev, ''Non-linear fluctuation-dissipation
relations and stochastic models in nonequilibrium thermodynamics'', Physica,
A 106, 443 (1981).

35. O.E.Lanford, ''Time evolution of large classical systems'', in
''Dynamical systems, theory and applications'', ed. J.Moser, Lectures Notes
in Physics, vol.38, 1975.

36. O.E.Lanford, ''On a derivation of Boltzmann equation'', in ''Nonlinear
phenomena. 1: The Boltzmann equation'', eds. J.L.Lebowitz, E.W.Montroll,
N.-H., Amsterdam, 1983.

37. H.Spohn, ''Theory of fluctuations and Boltzmann equation'', ibid.

38. G.N.Bochkov and Yu.E.Kuzovlev, ''Fluctuation-dissipation theory of
nonlinear viscosity'', Zh. Eksp. Teor. Fiz., 79, 2239, (1980) (transl. in
English in Sov.Phys.-JETP, No 12, 1980).

\end{document}